\documentclass[a4paper,11pt]{article}

\usepackage{graphicx}%
\usepackage{multirow}%
\usepackage{amsmath,amssymb,amsfonts}%
\usepackage{amsthm}%
\usepackage{mathrsfs, etoolbox}%
\usepackage{xcolor,authblk,float}%
\usepackage{hyperref, geometry}
\hypersetup{colorlinks,linkcolor={blue},citecolor={blue},urlcolor={black}}  

\newcommand{\der}[3]{\frac{d^#3 #1}{d#2^#3}}
		
		\newcommand{\pder}[3]{\frac{\partial ^#3 #1}{\partial #2^#3}}

\makeatletter 
\pretocmd\@bibitem{\color{black}\csname keycolor#1\endcsname}{}{\fail}
\newcommand\citecolor[1]{\@namedef{keycolor#1}{\color{red}}}
\makeatother
\begin{document}

\begin{center}
\baselineskip 24 pt 
{\LARGE \bf  
Quantum field theory at finite temperature for 3D periodic backgrounds }
\end{center}

\bigskip
\bigskip
\begin{center}

{\sc Luc\'ia Santamar\'ia-Sanz$^{1}$}

\medskip
{$^1$Department of Physics, University of Burgos, Plaza Misael Ba\~{n}uelos, Burgos, 09001, Spain}

\medskip
 
e-mail: {\href{mailto:lssanz@ubu.es}{lssanz@ubu.es}}

\end{center}

\medskip

\begin{abstract}
The one-loop quantum corrections to the internal energy of some lattices due to the quantum fluctuations of the scalar field of phonons are studied.  The band spectrum of the lattice is characterised in terms of the scattering data,  allowing to  compute the quantum vacuum interaction energy between nodes at zero temperature,  as well as the total Helmholtz free energy,  the entropy,  and the Casimir pressure between nodes at finite non-zero temperature.  Some examples of periodic potentials built from the repetition in one of the three spatial dimensions of the same punctual or compact supported potential are addressed: a stack of parallel plates constructed by positioning $\delta\delta'$-functions at the lattice nodes,  and an ``upside-down tiled roof" of parallel two-dimensional P\"oschl-Teller wells centred at the nodes.  They will be called \textit{generalised Dirac comb} and \textit{P\"oschl-Teller comb},  respectively.  Positive one-loop quantum corrections to the entropy appear for both combs at non-zero temperatures.  Moreover,  the Casimir force between the lattice nodes is always repulsive for both chains when non-trivial temperatures are considered, implying that the primitive cell increases its size due to the quantum interaction of the phonon field.  
\end{abstract}

\medskip
\medskip

\noindent

\section{Introduction}
\label{sec:intro}
Quantum field theory is one of the most successful theories in theoretical physics,  and constitutes a fruitful research area.  In particular,  the theory of complex scalar quantum fields living in classical backgrounds mimicking macroscopical objects has received much attention so far (see \cite{Mostepanenkobook, Miltonbook, Bordagbook} and references therein for a review).  A central concept to characterise these theories is the vacuum state,  in addition to the  $n$-point correlation function.   The Casimir effect \cite{Casimir1948,Spaarnay1957} is a physical manifestation of quantum fluctuations of the electromagnetic vacuum.  It allows the measure at zero temperature of quantum forces between closely spaced objects.  Such forces are due to changes in the mode spectrum of the quantum vacuum fluctuations with respect to the free space.  Not only the study of the zero-temperature quantum vacuum interaction energy (or Casimir energy) \cite{Casimir1948,  Spaarnay1957,  KK2008,  Kardar2008} in different geometries has frequently been addressed,  but also the thermal corrections to this value.  Although there exist many formalisms to tackle the problem,  the use of the boundary conditions allowed by the principles of quantum field theory to mimic bodies with arbitrary physical properties or topology changes in the space has been proved to be extremely useful \cite{Alvarez2015,  Asorey_2006,  Asorey2013}.  Some examples with possible application in condensed matter physics are given in \cite{Santamaria2019,Santamaria2020,Santamaria2020b}, where the authors novelly reinterpret the quantum system of a one dimensional chain as a one-parameter family of quantum pistons defined over the finite primitive cell interval,  by using suitable quantum boundary conditions.  In this way,  they were able to take advantage of some powerful tools of complex analysis and scattering theory to study the Casimir effect between the nodes of chains constructed as a repetition of potentials with compact support smaller than the lattice spacing.  However,  they only focused on 1+1 dimensional quantum field theories.  Now my aim is to generalise these results to 3+1 dimensional theories,  that more closely resemble real physical systems studied in materials physics or condensed matter laboratories.  Previous efforts in this direction can be found in \cite{Ines2016}, where the Casimir energy of a stack of parallel plates mimicked by pure $\delta$-plates at the points constituting a Cantor set has been studied by means of the Green's function formalism.  The resulting Casimir energy turns out to be positive,  involving a separation between the plates and a widening of the whole self-similar configuration.  However, notice that the positions of the plates in that work are given by the sequence $a, a/2,a/4,a/8...$, and not by equispaced positions in a chain with lattice spacing $a$. 

The objective of this work is to carry out a similar study but with regular lattices.  This paper presents a number of novelties not previously dealt with in the literature.  First of all,  I will address the problem of increasing the dimensions of the spacetime.  Indeed,  I compute the quantum vacuum interaction energy between nodes at zero temperature in lattices built from the repetition in one of the three spatial dimensions of the same punctual or compact supported potential.  In particular,  a stack of parallel plates constructed by positioning $\delta\delta'$-functions at the lattice nodes,  and an ``upside-down tiled roof" of parallel two-dimensional P\"oschl-Teller wells centred at the nodes will be studied.  They will be called \textit{generalised Dirac comb} and \textit{P\"oschl-Teller comb},  respectively.   Secondly,  I will generalise the technique firstly introduced in \cite{Donaire2020} to study the thermal correction to the vacuum energy between the plates and wells sitting at the lattice nodes in the two combs mentioned above, in order to determine how the lattice spacing is affected due to the thermal fluctuations of the fields.  In \cite{Donaire2020} the authors only analyse the thermal corrections to the total Helmhotlz free energy and one-loop quantum corrections to the entropy for a system of two isolated parallel plates,  and here I will do it for periodic backgrounds.  Finally,  it is worth stressing that these thermodynamic magnitudes in periodic systems will be derived in a convergent representation based on Cauchy integrals.  The major advantage of this method compared to others such as the one described in \cite{Ines2016} is that turning the integration contour towards the imaginary axis by a finite angle in the complex plane of frequencies avoids large oscillations of the Boltzmann factor,  and eases the numerical evaluation.  Furthermore,  it is easily applied to other periodic potential options provided that the scattering problem of one of the individual potentials forming the lattice alone in the the real line can be solved.  This means that the analytical method presented here is in a certain sense not model-dependent,  and this is a novel treatment of the problem. 

It is important to mention that the choice of the potentials that will be the building blocks of the periodic backgrounds is not arbitrary.  In fact that choice is based on finding several examples of potentials that are analytically exactly solvable, that are used in condensed matter physics,  and that have either a point support and a slightly larger one, in order to study the difference between both cases.  The generalised Dirac comb and the  P\"oschl-Teller one fit these characteristics perfectly.  In this paragraph I will delve into its many real-world uses in condensed matter physics.  On the one hand,  the first comb aforementioned is a generalisation of the \textit{Kronig-Penny model} in which the building block potential will be a combination of Dirac delta  potentials and its first derivative.  The \textit{Kronig-Penny model} is widely used in Solid State Physics to describe how an electron moves in a rectangular barrier-type lattice \cite{Kronig1931}.  The so-called pure \textit{Dirac comb} is a variation of the \textit{Kronig-Penney model} in which the rectangular barriers (or wells) transform into Dirac delta potentials with positive (or negative) coefficients. Dirac delta potentials are widely used as toy models for realistic materials like quantum wires \cite{Cervero2002}, and to analyse important physical phenomena such as Bose-Einstein condensation in periodic backgrounds \cite{BordagJPA2020},  or light propagation in 1D relativistic dielectric superlattices \cite{Halevi1999}. Despite being a rather simple idealisation of the real system, the $\delta$ function has been proved to correctly represent surface interactions in many models related to the Casimir effect \cite{Hennig1992, Fosco2009,Barton2004, Parashar2012}.  In addition, I will deal with a generalisation of this Dirac comb by introducing the first derivative of the Dirac delta place at the same point as the Dirac delta,  in order for the whole potential to be regularised \cite{Gadella2009}.   The first derivative of the delta potential has been used by M. Bordag in the study of monoatomically thin polarisable plates formed by lattices of dipoles \cite{Bordagprima2014}, and to study resonances in 1D oscillators \cite{Alvarez2013}, to mention just a couple of applications.  In fact,  the $\delta'$ potential has been frequently used in physics since the early 20th century  as an approximation of short range potentials (see \cite{Albeverio1984} and references therein),  and its inclusion significantly modifies the energy spectrum of the problems considered \cite{Albeverio2013}.  To sum up,  the \textit{generalised Dirac comb} to be used in this work is built as a repetition of the following potential with punctual support smaller than the lattice spacing $a$:
\begin{equation}\label{Diracpot}
V_{\delta\delta'}(z)=  w_0 \delta(z)+2w_1 \delta'(z), \qquad w_0, w_1 \in \mathbb{R}.
\end{equation}
On the other hand, the \textit{P\"oschl-Teller comb} or PT comb (where the potential with compact support is related to the sine-Gordon kink \cite{Guilarte2011}) will be studied. The P\"oschl-Teller potential \cite{PTeller1933, Negro2016} is known to have application in a variety of areas in physics such as astrophysics, quantum many-body systems or supersymmetric quantum mechanics,  but also in condensed matter. For instance, it has been employed in \cite{Park2015} to numerically represent potential barriers in bilayer graphene. In \cite{Hartmann2017} the relativistic one-dimensional P\"oschl-Teller potential problem has been studied to describe graphene waveguides. Another promising example is given in \cite{Tomak2005},  where quantum wells\footnote{Quantum wells are structures consisting of alternating thin layers of semiconductors with different band-gaps. They can confine particles in the dimension perpendicular to the layer surface, whereas the movement in the other dimensions is not restricted. The concept was proposed by  H. Kroemer \cite{Kroemer1963} and by Z. Alferov and R.F. Kazarinov \cite{Alferov1963} in the mid-1960s.} mimicked by P\"oschl-Teller confining potentials are considered to calculate intersubband absorption and other non-linear optical properties. It seems to be possible to create this type of quantum wells with the recent nanotechnology progress \cite{Radovanovic2000} thus it seems necessary to have an in-depth knowledge of their properties from a theoretical point of view as well.   The \textit{P\"oschl-Teller comb} to be considered in this work is built as a repetition of the following potential with compact support $\epsilon$ smaller than the lattice spacing $a$:
\begin{equation}\label{PTpot}
V_{PT}(z)= -\frac{2 \Theta(-z+\epsilon/2)\Theta(z+\epsilon/2)}{\cosh^2(z)},
\end{equation}
being $\Theta(z)$ the  Heaviside function. 

The paper is organised as follows.  In Section 2,  the necessary already existing formulas for determining the band spectrum of generic periodic potentials are collected,  and then particularised for the P\"oschl-Teller  and the generalised Dirac combs.  In Section 3,  the quantum vacuum interaction energy for scalar fields propagating under the influence of such three-dimensional periodic backgrounds  is computed,  and then particularised for the two cases of interest.  In Section 4,  I derive a general representation of the free energy, entropy and Casimir pressure between nodes for arbitrary temperature,  and I compute them numerically for the two particular lattices mentioned above.  Finally,   the conclusions and open questions are presented in Section 5.  Throughout the paper the system of natural units will be used,  i.e.  $\hbar=c=k_B=1$ .

\section{Band spectra}
\label{sec:spectra}

In this section  the spectrum of the quantum vacuum fluctuations of the phonon field interacting with the generalised Dirac comb,  as well as with the P\"oschl-Teller comb,  will be studied.  Notice that one can pass from a $N$-particle description of a lattice to a classical field theory \cite{Simons},   and study the universal features and experimental observables of the model.  The classical solutions of the equations of motion in such a field theory are lattice vibrations propagating as sound waves. This collective behaviour of the lattice atoms is a quasiparticle called phonon.  In this way,  the solid is treated as if contains different weaking interacting particles (or phonons) in the vacuum.  So here I will only study weak effects of lattice vibrations on the solid (i.e.   small displacements of the atoms with respect to the equilibrium positions),  because I want to preserve its integrity.   If one then studies the small fluctuations around a classical solution of the action up to second order,  and treats the fluctuation as a quantum field,  the Hamiltonian which describes the whole lattice becomes that of an infinite collection of non-interacting harmonic oscillators.  Each energy level of the harmonic oscillator is an accumulation of $n_k$ elementary entities or quasiparticles,  each one having energies $\omega_k$.  These quasiparticles are characterised only by their energy,  and not by any other quantum number.  Consequently,  they are bosons because one can create a state with any number of identical excitations by repeatedly applying creator operators on the vacuum state.  For that,  it is often said that the vibrational modes of the solid can be identified with bosonic quasiparticles that can be studied within a quantum scalar field theory.  This will be the starting point of this work.  Let us first analyse the quantum mechanical problem to determine the spectrum of the fluctuation modes in the lattice.

\begin{figure}[H]
\centering
\includegraphics[scale=0.12]{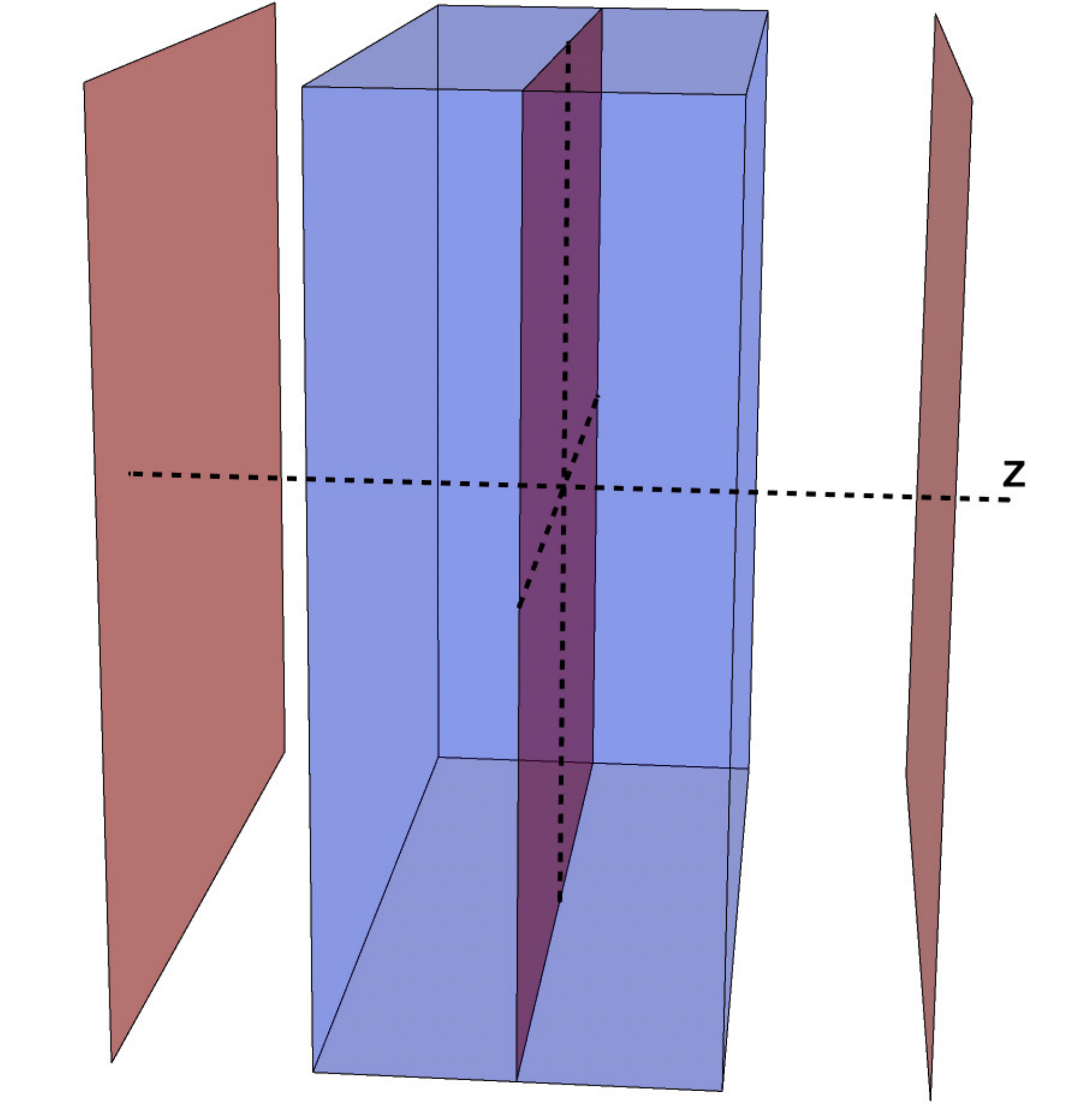}\qquad \qquad \qquad  \includegraphics[scale=0.22]{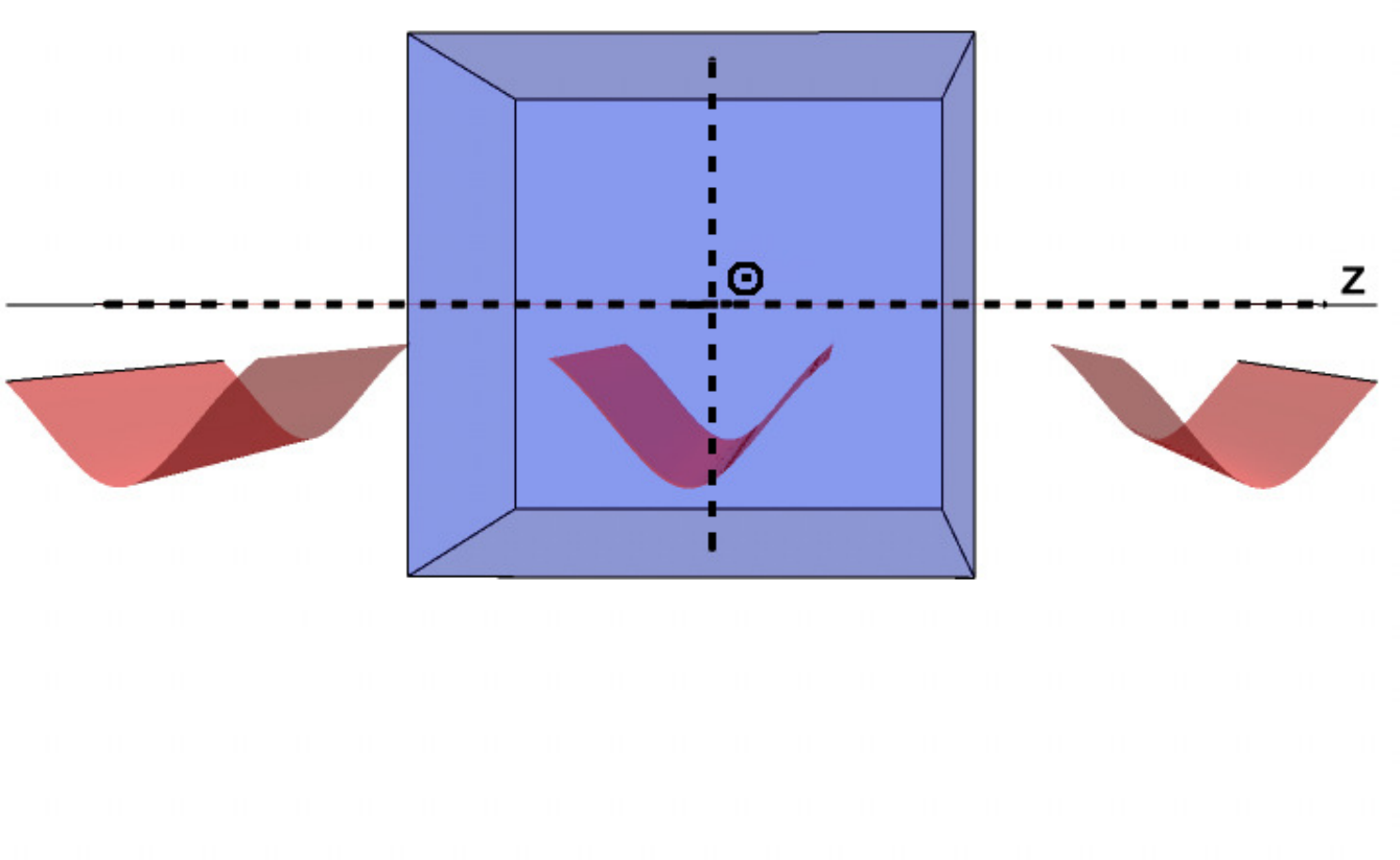}
\caption{\footnotesize Left: Lattice of parallel Dirac $\delta \delta'$ plates (red) in three spatial dimensions and primitive cell (blue). Right: Lattice of truncated PT potentials (red) in three spatial dimensions and primitive cell (blue).} 
\label{fig:highdim}
\end{figure}

The potential representing the three dimensional combs under consideration is given by:
\begin{eqnarray}
&&U(\vec{x})= \sum_{n\in \mathbb{Z}} V(z-na), \qquad \textrm{being} \qquad V(z) \left\{
	       \begin{array}{ll}
		 \neq 0      & \mathrm{if\ } |z|\leq \epsilon{\color{black} /2},  \\
		 = 0 & \mathrm{if\ } |z|> \epsilon{\color{black} /2},
	       \end{array}
	     \right. \label{chainpotential3D}
\end{eqnarray}
and $\vec{x}=(\vec{x_\parallel},z)$, with $\vec{x_\parallel} \in \mathbb{R}^2$.  Notice that in the $z$ direction the potential of the comb is built from the repetition of another one-dimensional potential $V(z)$, which stands at the lattice nodes $na$ and vanishes outside small intervals of the form $[na-\epsilon/2, na+\epsilon/2]$ included in the unit cell $[na-a/2, na+a/2]$ with $a\geq\epsilon>0$. For instance,  the geometry either for the generalised Dirac comb and the P\"oschl-Teller one is represented in Figure \ref{fig:highdim}.  All figures and graphs in this paper have been produced with the software  \textit{Mathematica}  from Wolfram Research.
  In the direction orthogonal to the plates, i.e. (0,0,1),  a problem analogous to that of a one-dimensional comb presented in \cite{Santamaria2019,  Santamaria2020,  Santamaria2020b} appears.  However, in the two directions parallel to the plate, (1,0,0)  and (0,1,0), since the plates or the P\"oschl-Teller wells are thought to have infinite area,  the scalar field moves freely without boundaries and consequently, one recovers the spectrum of a free particle in these two directions.  The frequency of the field modes in the whole system would thus be\footnote{Notice that $\vec{k}=(\vec{k_\parallel},k)$, with $\vec{k_\parallel} \in \mathbb{R}^2$. } $\omega=\sqrt{\vec{k_\parallel}^2+k^2}$,  and one could focus on solving the problem the spectrum in the $z$ direction.  Notice that if the plates or wells had a finite area and some boundary conditions for the wave function of the scalar field were imposed on the $x, y$ axes (representing a more realistic material), the frequencies on the parallel dimensions would be quantised. This type of cases are left for future further investigation.  

In \cite{Santamaria2020b} the authors reinterpret the quantum system of the comb by using general quantum boundary conditions,  and following the formalism described in \cite{Asorey2013}.  They explain that such a comb in one-dimensional spaces can be interpreted as a one-parameter family of quantum pistons,  where the middle piston membrane is represented by the individual potential $V(z)$ placed at the middle point of the primitive cell, and the extremal points of the primitive cell correspond to the external walls of the piston placed at $z=na\pm a/2$.   At these walls some specific boundary conditions must be imposed over the fields.  Here I will follow this prescription.  Notice that the quantum scalar field $\psi(z)$ satisfies the Floquet-Bloch boundary conditions at the extremal points of the lattice cell
\begin{eqnarray}\label{Bloch}
\psi_\theta(z+a)=e^{-i\theta} \psi_\theta(z), \qquad \qquad \partial_y\psi_\theta(y)|_{y=z+a}=e^{-i\theta}  \,\partial_y\psi_\theta(y)|_{y=z} ,
\end{eqnarray}
where $\theta=-q a$ is related to the quasi-momentum $q$ of the Bloch wave in the first Brillouin zone\footnote{The first Brillouin zone is the locus of points in the reciprocal lattice that are closer to its origin than they are to another point in the lattice. It is the primitive cell in the reciprocal space.}.  In quantum mechanics,  the dimensionless time independent (i.e. after Fourier transform) Schr\"odinger equation for a quantum scalar field in the background of a one-dimensional comb is:
\begin{equation}
\hat{K}_{comb}\, \,  \phi_\omega(z) \equiv\left( -\pder{}{z}{2} + {\color{black} U}(z)\right) \phi_\omega(z)= \omega^2 \phi_\omega(z),
\end{equation}
where $\omega=k$ are the frequencies of the quantum field modes\footnote{In QFT, the energies of the set of one-particle states are given by the squared root of the eigenvalues of the non-relativistic Schr\"odinger  operator $\hat{K}$ whereas in quantum mechanics the frequencies of the quantum field modes are given by the eigenvalues of $\hat{K}$. Most references tend to mix both notations.}. They are related to the spectrum of the one-particle states of the Schr\"odinger operator since  $\omega^2 \in \sigma(-\partial_z^2+U(z))$.  The band spectrum of the lattice can be written in terms of {\color{black}the transmission amplitude $t(k)$, and the reflection amplitudes for incoming particles from the left $r_R(k)$ and in the opposite direction $r_L(k)$ for the scattering problem related to the Schr\"odinger operator of the compact supported potential from which the comb is built:
\begin{equation}\label{Hindiv}
\hat{H}_V = -\der{}{z}{2} + V(z).
\end{equation}
It is worth clarifying that talking about scattering for the potential $U(z)$ is meaningless because the scattering solutions can be defined only for potentials with compact support which reduce to zero for asymptotic values of the coordinate, and $U(z)$ is not among them.  Nevertheless, this is exactly what happens if one considers only one potential $V(z)$, from which the comb is built, alone in the whole real line.  Scattering here does make sense.  In this way,  the spectral equation is \cite{Santamaria2020b}
\begin{eqnarray}\label{spectralfcomb}
&&\hspace{-7pt} f_{\theta}(k) \equiv  \cos(\theta) - h_V(k)=0,\, \,\,  \textrm{with} \,\,\, h_V(k)= \frac{1}{2t(k)} \left[ e^{-ika} + e^{ika} (t^2(k)-r_R(k)r_L(k))\right],\nonumber\\
&&\hspace{-7pt}\textrm{and} \,\,\,  \theta \in [-\pi/\pi].
\end{eqnarray}
The real solutions of the spectral equation determine the energy levels of the crystal. Hence, the band spectrum arises as the different branches $E(q)$ of the equation $f_{-qa}(\sqrt{E})=0$ for every value of $q\in[-\pi/a, \pi/a]$.  Since $\cos(\theta)$ in eq.  (\ref{spectralfcomb}) is a bounded function and consequently 
\begin{equation}|h_V(k)| \leq 1,
\end{equation} the energy spectrum of the system is organised into allowed and forbidden energy bands and gaps. Additionally, the discrete set of wave vectors characterised by $\{k_i \in \mathbb{R}\, /\, |h_V(k_i)| = 1\}$ determines the lower and higher values of $k$ for each allowed band. 

The spectrum of positive energy bands contains an infinite number of allowed bands, since an effective model of an  infinite linear chain is considered.  However, it is worth stressing that the possible solutions of eq.  \eqref{spectralfcomb} for imaginary momenta ($k=i\kappa$, with $\kappa>0$) so that the energy is negative ($E=-\kappa^2$) form negative energy bands. Again, the allowed energies for these bands satisfy $|h_V(k=i\kappa)| \leq 1$.

All the characterisation of the spectrum described in this section applies for any comb built as a repetition of a potential with compact support smaller than the lattice spacing sitting at the lattice nodes.

\subsection{Generalised Dirac comb}
The secular equation and the scattering data for $V_{\delta\delta'}$ \eqref{Diracpot}, i.e.
\begin{eqnarray}\label{secfunctDirac}
f_q^{Dc}(\Omega, \gamma, k,a )&\equiv &\Omega \cos( q a) +\cos (k a)+\frac{\gamma}{2} \frac{\sin (k a)}{k}=0,\nonumber\\
t(k)&=& \frac{-2 k\Omega}{i\gamma+2k}, \nonumber\\
r_R(k)&=& \frac{-i\gamma-2k \sqrt{1-\Omega^2}}{i\gamma+2k}, \nonumber\\
r_L(k)&=& \frac{-i\gamma+2k \sqrt{1-\Omega^2}}{i\gamma+2k},
\end{eqnarray}
with  $\gamma= w_0/(1+w_1^2)$ and $\Omega = (w_1^2-1)/(w_1^2+1)$,  were previously studied in \cite{Santamaria2020b}.  The main results are collected here to completeness, and ir order to compare them with the  P\"oschl-Teller case,  which has not been correctly studied before in \cite{Santamaria2020b}. 

Once the spectral equation is solved numerically with \textit{Mathematica}, one can evaluate the energy of the allowed bands by means of $E=k^2$. Two different cases arise depending on the values of the couplings $w_0,w_1$: situations in which there is a negative energy band of localised states (because if $w_0<0$ the $\delta$-$\delta'$ potential admits one bound state) and occasions in which the lowest energy band is positive and the phonons propagate freely along the crystal as plane waves. Figure \ref{fig:spectrumDirac} allows the comparison between the first two allowed energy bands for the pure Dirac comb ($w_1=0$) and for the generalised Dirac comb.
\begin{figure}[h]
\centering
\includegraphics[width=0.35\textwidth]{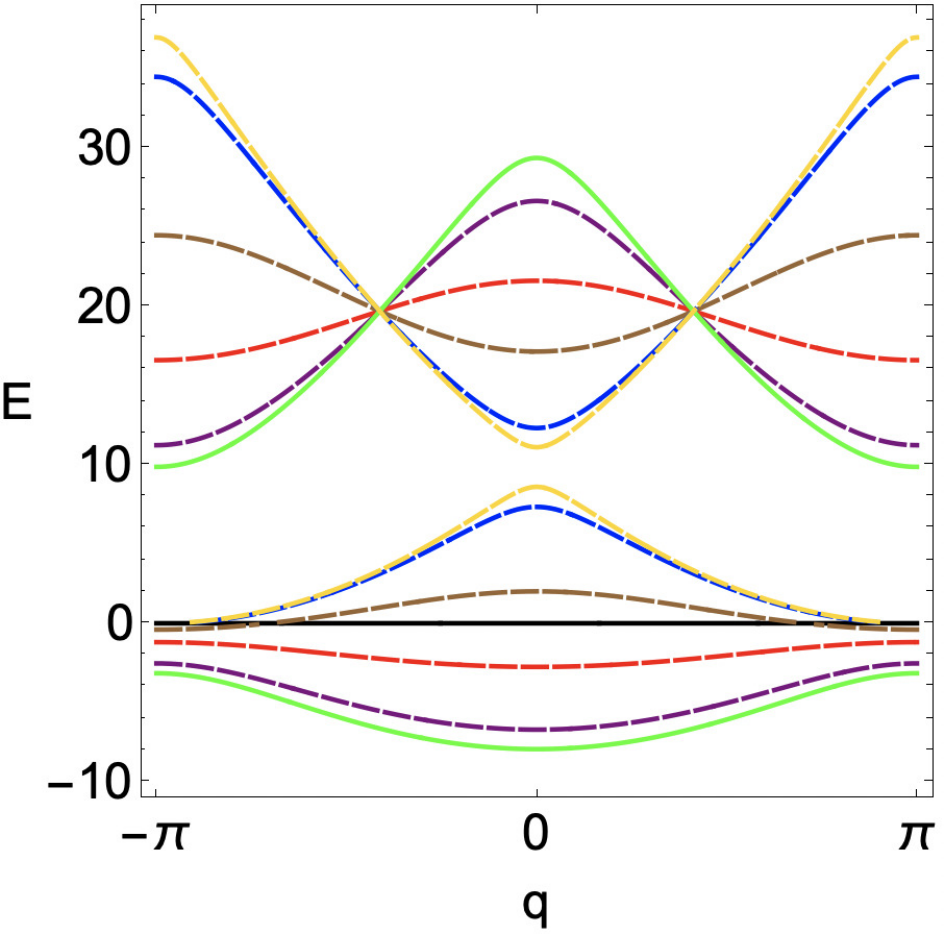}
\qquad \qquad \qquad 
\includegraphics[width=0.338\textwidth]{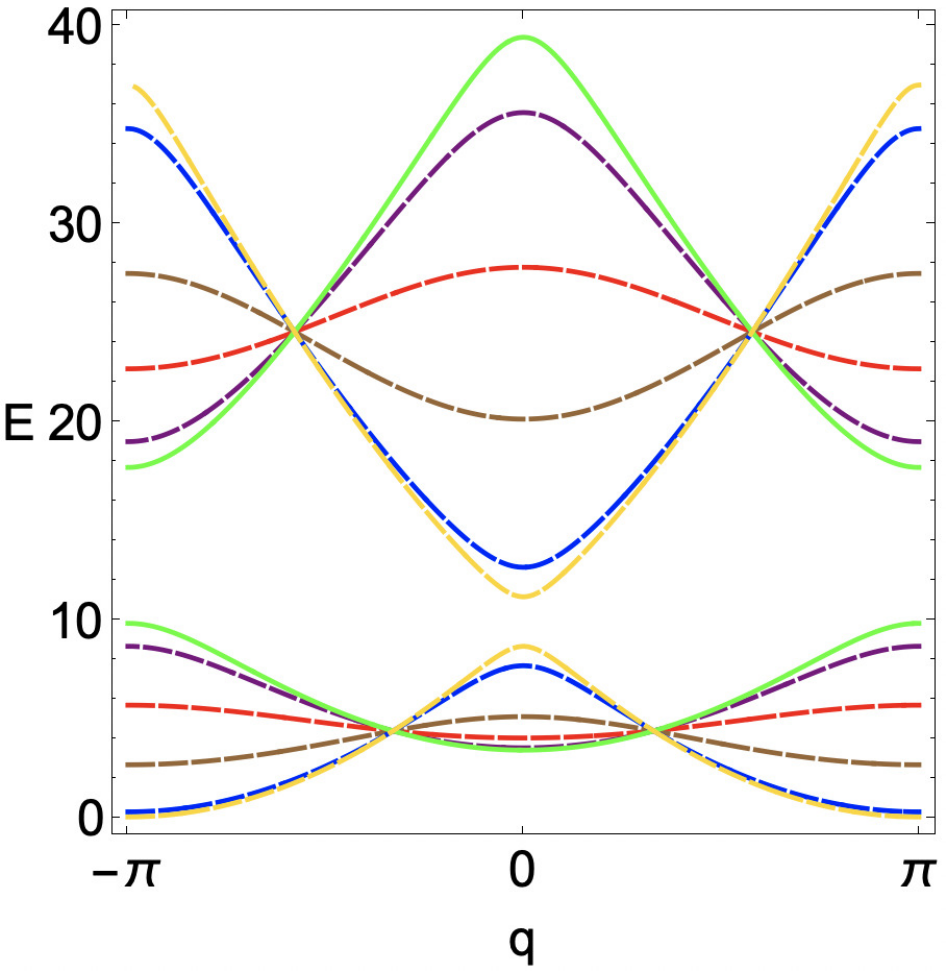}
\caption[\footnotesize First two allowed energy bands for the pure Dirac comb and the generalised Dirac comb for different values of $w_0,w_1$]{\footnotesize First two allowed energy bands for the pure Dirac comb (solid green curve) and the generalised Dirac comb (dashed lines). On the left $w_0=-5$ for all the cases and on the right $w_0=5$. In both plots $w_1=0.3$ (purple), $w_1=0.75$ (red), $w_1=1.5$ (brown), $w_1=5$ (blue) and $w_1=-10$ (yellow). The black line in the  plot on the left represents the zero energy level.  In both plots $a=1$, $E=k^2$ and $q$ is the quasi-momentum in the first Brillouin zone.}
\label{fig:spectrumDirac}
\end{figure}

It is well-known that for spin 1/2 fermionic carriers in crystals, depending on the existence or not of a gap between the negative energy band and the first positive one, the charge carriers could spontaneously go or not from localised states of negative energy to propagating states of positive energy. The first situation corresponds to a conductor behaviour if the lowest energy band is not completely filled. The second case above described represents a semiconductor or insulator (depending on the size of the gap) whenever the  valence band is completely filled. Notice that when there are many fermionic charge carriers, the Dirac-Fermi statistics must be introduced in the analysis. However, in the problem present here, only scalar fields and their vibrational spectrum are involved.  Due to the Bose-Einstein statistic, all the modes of the spectrum are available to be occupied by bosons without any restriction as to their number. One cannot talk about valence and conducting bands separated by a gap of prohibited energy because the way in which modes are occupied is completely different from the fermionic case. What really happens in the bosonic case is that certain frequencies are forbidden for phonon propagation. This is how these band spectra are to be interpreted in this work.

\subsection{P\"oschl-Teller comb}
Following \cite{Guilarte2011},  the scattering coefficients for $V_{PT}(z)$ are
\begin{eqnarray}\label{scatkink}
r(k)\!&=&\! \frac{e^{i\epsilon k}\Lambda(\Lambda + 2k (k+i\tanh (\epsilon /2)))}{\Delta(k)}
- \frac{e^{-i\epsilon k}\Lambda(\Lambda + 2k (k-i\tanh (\epsilon /2)))}{\Delta(k)},\nonumber\\
t(k)\!&=&\!\frac{4k^2(k^2+1)}{\Delta(k)},\nonumber\\
\Delta (k) \!&=&\!-e^{2i\epsilon k} \Lambda^2 +\! [\Lambda+2k(k-i\tanh (\epsilon /2))]^2,
\end{eqnarray}
with $\Lambda=1-\tanh^2 (\epsilon/2)$. Notice that the reflection coefficients fulfil  $r_R(k)=r_L(k)=r(k)$ due to the parity symmetry of the potential. The poles $k=i\kappa$ with $\kappa>0$ of the determinant of the scattering matrix  ($\textrm{det} \, S= t^2-r_Rr_L$) are the bound states of the kink-comb spectrum.  In this comb, the well known state of the non-spatially truncated kink potential (i.e.  eq. \eqref{PTpot} without Heaviside functions) with $k=i$ is not a bound state. 

Once more, the allowed bands are determined by the real solutions of the spectral equation (obtained by replacing \eqref{scatkink} into \eqref{spectralfcomb}), which takes the form
\begin{eqnarray}\label{fsecPTcomb}
&&\hspace{-7pt}f^{PTc}_q(a,k, \epsilon)\equiv\cos(q a) - \frac{\Sigma \cos(ka)-\Upsilon \sin(ka)}{k^2(k^2+1)(1+\cosh \epsilon)}=0,\qquad \textrm{with}\nonumber\\
&&\hspace{-7pt}\Upsilon= 2k \tanh \frac{\epsilon}{2}(1+k^2+k^2 \cosh \epsilon) +  \Lambda \cos(k\epsilon) \sin(k \epsilon), \nonumber\\
 &&\hspace{-7pt}\Sigma= k^2(3+k^2)+k^2(-1+k^2)\cosh \epsilon + \Lambda \sin^2(k\epsilon).
\end{eqnarray}
In Figure \ref{fig:spectrumPT} the two lowest energy bands of the P\"oschl-Teller comb are shown for different values of the unit cell size.
\begin{figure}[H]
\centering
\includegraphics[width=0.357\textwidth]{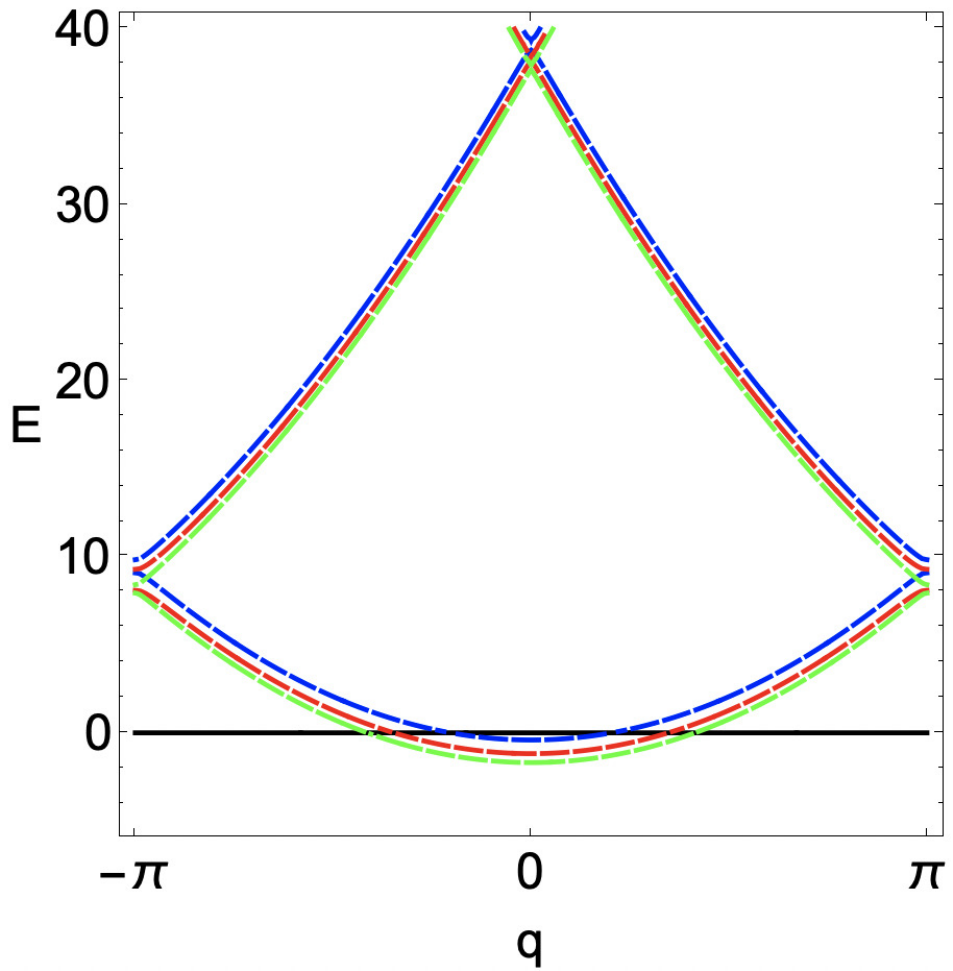} \qquad \qquad \qquad \includegraphics[width=0.36\textwidth]{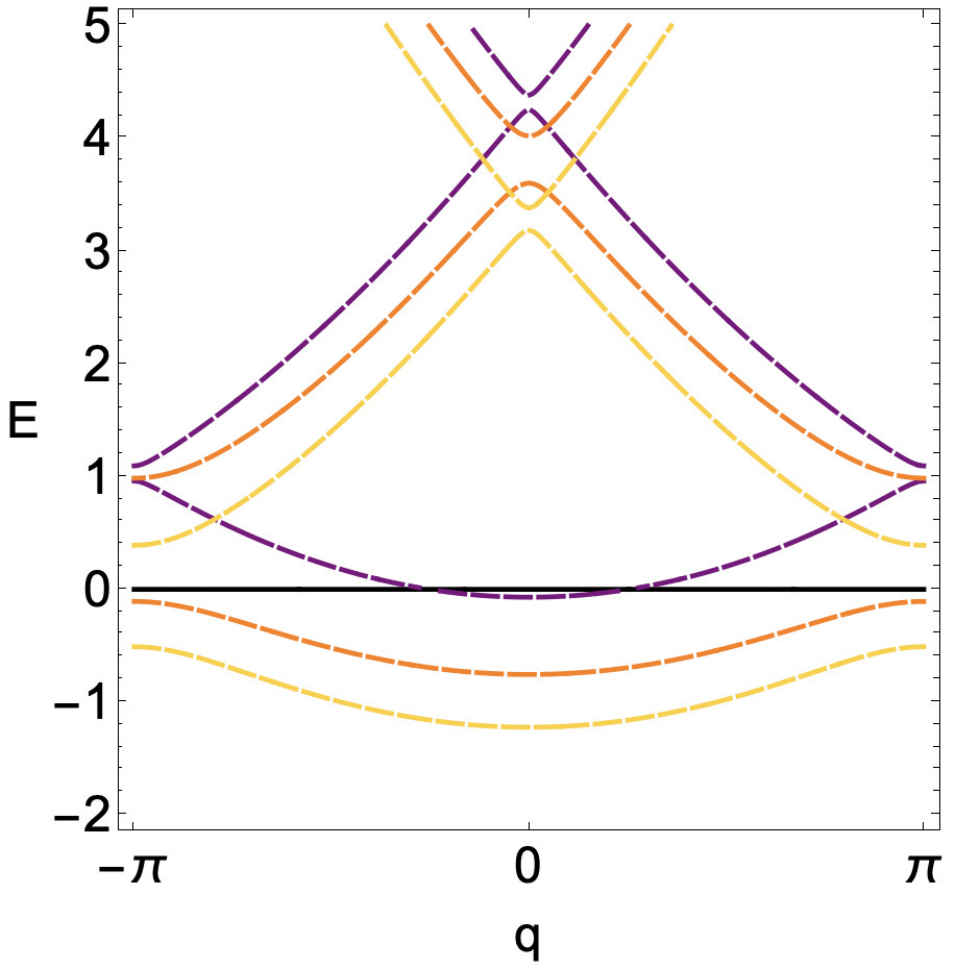} 
 \caption[]{\footnotesize Left: First two allowed energy bands for the P\"oschl-Teller comb for $a=1$ and different values of the compact support $\epsilon=0.2$ (blue), $\epsilon=0.6$ (red) and $\epsilon=0.9$ (green).  Right: First two allowed bands for  $a=3$ for compact supports $\epsilon=0.1$ (purple), $\epsilon=1$ (orange) and $\epsilon=2.5$ (yellow). In these plots $E=k^2$ and $q$ is the quasi-momentum in the first Brillouin zone.}
\label{fig:spectrumPT}
\end{figure}
As there can be seen in Figure \ref{fig:spectrumPT}, whenever  $a>\epsilon>0$, there is gap between the maximum value of the energy in the first band and the minimum of the second band. Moreover, from the graph on the top in Figure \ref{fig:spectrumPT} it is clear that the kink comb always has a band with negative energy. This comes from the fact that the potential \eqref{PTpot} always has one bound state with momentum $k=i\kappa$, being $0<\kappa<1$ for $a=1$. If $a$ increases, there is still a negative energy band for any value of the compact support $\epsilon$, as shown in the plot on the bottom in Figure \ref{fig:spectrumPT}.  It will be crucial to take this fact into account when calculating the vacuum energy of this system in the corresponding QFT, because whenever there is a negative energy band, the QFT ceases to be unitary and absorption phenomena in quantum wires appear at zero and non-zero temperature.

\section{Vacuum energy at zero temperature}
\label{sec:vacuum}
Once the non-relativistic quantum mechanical problem has been solved, one could perform the second quantisation  to focus on computing some relevant magnitudes of the associated QFT.  In the periodic structure (\ref{chainpotential3D}),  the quantum vacuum interaction energy $E_0$ physically gives the phonon contribution to the internal pressure of the chain if one interprets the quantum scalar field as phonons in the crystal. Hence, $E_0$ enables to study the dilatation or reduction of the primitive cells in the lattice due to the fluctuations of the phonons.

In \cite{Santamaria2019} the quantum vacuum energy per unit cell of a one-dimensional  chain is computed by using the spectral zeta function.  The authors explain that  the zeta function of a comb is the continuous sum of zeta functions over the dual primitive cell of Bloch quasi-momenta.  Hence,  the sum over the spectrum of the comb can be done by performing the summation over the quasi-momentum of all the discrete spectra characterised by $f_\theta(k)=0$ that arises for each value of the parameter $\theta$. That is,  the finite quantum vacuum energy of the comb, $E_0^{comb}$, can be obtained from the finite quantum vacuum energy $E_0^V$ of the quantum scalar field confined between two plates placed at $z=\pm a/2$ represented by the boundary condition associated to \eqref{Bloch}, and under the influence of the individual potential  that forms the comb $V(z)$\eqref{chainpotential3D}  placed at $z=0$, viz. :
\begin{eqnarray}\label{spectralE0bis}
&&\hspace{-25pt}E_0^{comb}= \frac{1}{2} \sum_{\omega^2 \in\sigma(\hat{K}_{comb})} \!\!\! \omega= \frac{1}{2\pi} \int_{-\pi}^{\pi} d\theta \, E_0^{V}(\theta).
\end{eqnarray}

It is necessary at this point to generalise the formalism introduced in  \cite{Santamaria2019} in order to apply it for combs in three dimensions,  and also for combs with negative energy bands, which was not the case presented in the aforementioned paper.    When bound states exist in the spectra of the corresponding quantum mechanical problem, there is a part of the spectrum of fluctuations that corresponds to negative energies. This is not a problem at all in quantum mechanics but an obstacle when considering QFT. In quantum mechanics the preservation of unitarity implies that the Hamiltonian operator for a system with boundaries must be a self-adjoint operator. Otherwise, there will exist a non-trivial probability flow across the boundary. In QFT one treats with a grand canonical ensemble of particles. Preservation of unitarity in QFT implies that the Hamiltonian operator must be a non-negative self-adjoint operator. It must be self-adjoint in order for the eigenvalues to be real numbers and hence measureables. Furthermore, despite the case of quantum mechanics, in QFT the Hamiltonian must also be non-negative. This is due to the fact that the presence of negative energy states in the Schr\"odinger problem that gives the one particle states of QFT results in an imaginary quantum vacuum energy caused by the absorption and emission of the scalar field fluctuations by the plates. This effect is the bosonic version of the Schwinger effect \cite{PhysRev.82.664}. A way to solve this problem is the introduction of a mass for the quantum fluctuations. In this way, the whole energy spectrum is pushed towards values greater than or equal to zero, and the unitarity of the QFT is guaranteed.  Mass fluctuations are of more than just academic importance, and it could be generated in a real experiment via crystal symmetry breaking \cite{ZeljkovicNature2015, OkadaScience2013, WangNature2016} in fermionic systems,  for instance.  Remember that phonons are quasiparticles describing the collective excitations of atoms in lattices.  Crystals which contain more than one type of atom, and elemental crystals with complex structures,  may have optical phonons whose energy gap when the wavevector is zero is interpreted as a mass in a Klein-Gordon scalar field theory.   This means that their behaviour in the material can be explained if such phonons are though of as free particles with an effective mass. There are many applications of massive phonons in condensed matter physics and astrophysics \cite{Baggioli_2020,  Visser:2004qp,  Marino2019}.  In that sense,  it should not be strange to  assign masses to the phonons,  since we are not considering acoustic phonons, which are Goldstone bosons related to the translational symmetry in the lattice, but optical ones.  Back to the problem at hand,  $\omega=\sqrt{\vec{k}_\parallel^2+k^2+m^2}$ will be the frequencies of the quantum field modes from now on.   The Casimir energy per unit area of the plates/wells is the summation over the frequencies of the field modes spectrum:
\begin{eqnarray}
&&\hspace{-15pt}\frac{E_0^{comb}}{A}=  \frac{1}{2\pi}\!\! \int_{-\pi}^{\pi} \!\!d\theta\!\! \underset{k \in Z(f_\theta)}{\sum} \!\int_{\mathbb{R}^2} \!\!\frac{d\vec{k}_\parallel}{(2\pi)^2} \sqrt{m^2+k^2+\vec{k}_\parallel^2}.
\end{eqnarray}
This result is ultraviolet divergent due to the contribution of the energy density of the free theory in the bulk and the self-energy of the plates.  Firstly,  to perform the integral over the parallel momenta,   it is necessary to introduce a regulator.  For instance, one could introduce an exponentially decaying function and perform the integration over the parallel modes:
\begin{eqnarray}
 &&\hspace{-10pt} \lim_{\epsilon \to 0}\,\, \underset{k}{\sum}  \,\, \int_{\mathbb{R}^2} \frac{d\vec{k}_\parallel}{(2\pi)^2} \sqrt{m^2+k^2+\vec{k}_\parallel^2} \, e^{-\epsilon \vec{k}_\parallel^2}=\lim_{\epsilon \to 0} \,\, \,\, \underset{k}{\sum} \,\,  \int_{0}^\infty \frac{dk_\parallel}{2\pi} k_\parallel \, \sqrt{m^2+k^2+k_\parallel^2} \, e^{-\epsilon k_\parallel^2}  \nonumber\\
&&\hspace{-10pt}  =\lim_{\epsilon \to 0}  \underset{k}{\sum}  \frac{1}{2\pi}\left( \frac{\sqrt{\pi}}{4 \epsilon^{3/2}} +  \frac{\sqrt{\pi} (m^2+k^2)}{4 \sqrt{\epsilon}} - \frac{(m^2+k^2)^{3/2}}{3}  +o(\sqrt{\epsilon})\right) = -  \underset{k}{\sum} \frac{(m^2+k^2)^{3/2}}{6\pi}.
\end{eqnarray}

Notice that in the final step, the terms proportional to $\epsilon^{-3/2}$ and $\epsilon^{-1/2} $ have been removed to eliminate the contribution of the parallel modes to the dominant and subdominant divergences,  respectively.  Dominant or subdominant divergence refers to the degree of ultraviolet divergence.  For the system of a scalar field confined between two plates, the dominant divergence is a term proportional to the energy raised to the power $(D+1)/2$ with $D$ being the spatial dimension of the theory,  and the subdominant as the energy raised to $D/2$ \cite{Asorey2013}.  Both terms are divergent in the ultraviolet regime. Whilst the former is associated to the density energy of the free theory in the bulk, the latter is due to the self-energy of infinite area plates. This classification of the divergences will also be used here from now on.  

Up to this point one could rewrite the Casimir energy as
\begin{eqnarray}\label{eqgen5}
&&\hspace{-20pt}\frac{E_0^{comb}}{A}=-  \frac{1}{2\pi} \int_{-\pi}^{\pi} d\theta \,  \,\, \underset{k \in Z(f_\theta)}{\sum} \frac{(m^2+k^2)^{3/2}}{6\pi}.
\end{eqnarray}

Now that one is only working with the orthogonal direction to the plates,   the summation over the spectrum of the comb must be splitted into two terms: the one corresponding to the firsts bands or the part of them which corresponds to  states with negative energy, and the other related to all the positive energy bands, either the rest of the first bands and all the upper bands, as the latter only contain positive energy states (see for instance figure \ref{fig:conteo}). 
\begin{figure}[h]
\centering
\includegraphics[width=0.4\textwidth]{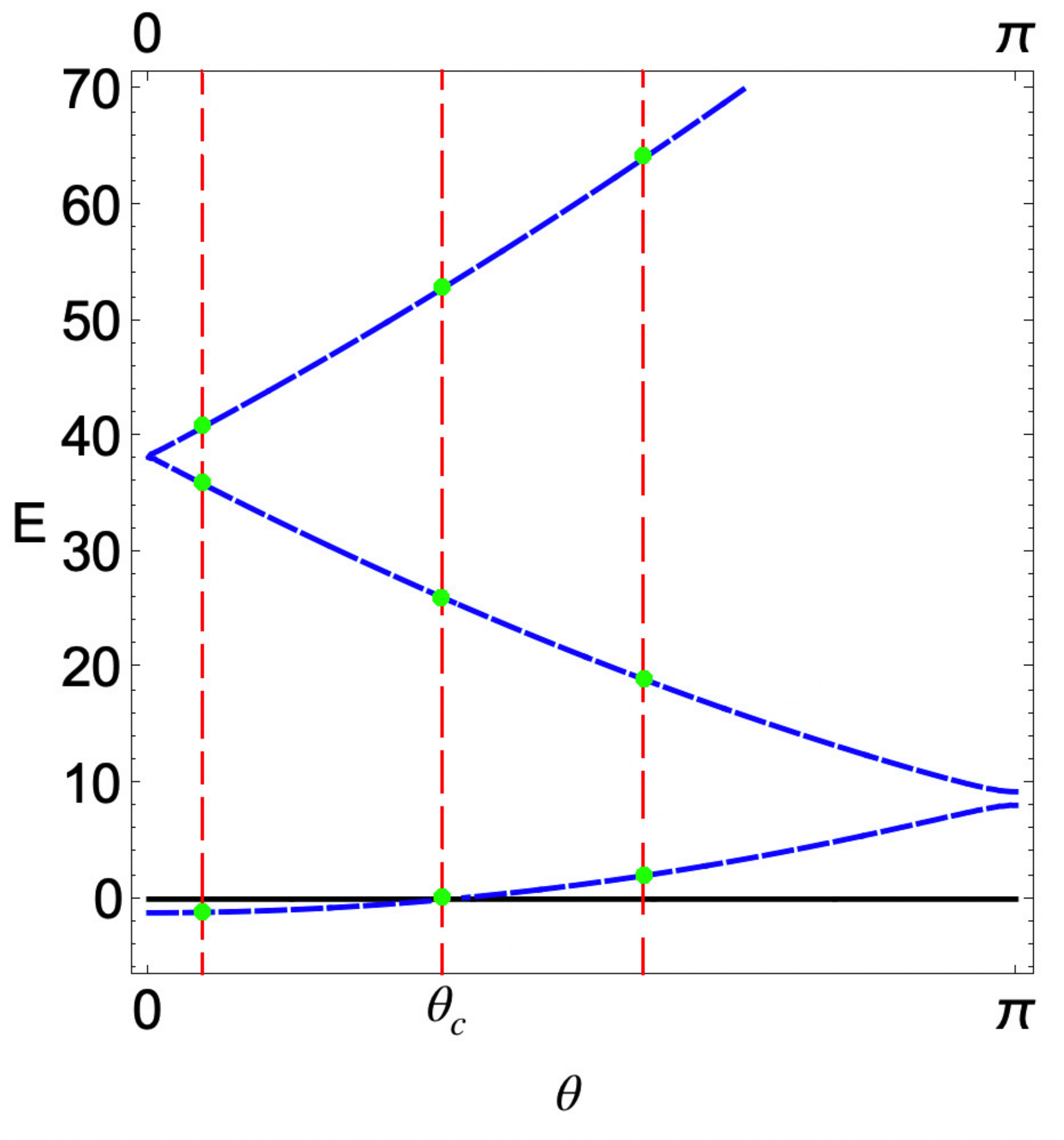}
\caption[\footnotesize Graphical explanation of how to perform frequency counting to calculate the quantum vacuum interaction energy in the PT comb.]{\footnotesize First three bands of the spectrum for the PT comb characterised by $a=1, \epsilon=0.6$. For each fixed value of the parameter $\theta \in [0, \pi]$, one has to sum over the wave vector $k$ or $\kappa$ associated to the states highlighted by green dots. Notice that in quantum mechanics $E=k^2$ when $E>0$ and $E=(i\kappa)^2$ when $E<0$.  In these plots, $\theta=q a= q$ is the quasi-momentum in the first Brillouin zone.}
\label{fig:conteo}
\end{figure}

Regarding for instance the PT comb,  the first band contains states with negative energy. Consequently, one has to sum up the momentum of the bound state with $k=i\kappa, \, \kappa>0$ and then integrate over the quasi-momentum. But instead of considering the whole Brillouin zone, one has to integrate  only over the interval $[-\theta_c, \theta_c]$. $\theta_c$ is related to the value of the quasi-momentum in the first Brillouin zone beyond which there are no negative energies in the first band of the comb spectrum (see figure \ref{fig:conteo}). It could be obtained from the secular equation as:
\begin{equation}
\cos \theta_c - h_V(k=0)=0.
\end{equation}
Notice that $k=0$ for $\theta=\theta_c$. If all the first band is a negative  energy band, then $\theta_c=\pi$. On the other hand, if $\theta_c< \pi$,  in the interval $\theta \in (\theta_c, \pi]$ the first band only includes states with positive energy. 

Furthermore,  the mass which should be introduced in the theory must satisfy the condition $m\geq \kappa$ being $\kappa$ the wave vector of the  negative energy bound state of the spectrum  for a given $\theta<\theta_c$. In this case, $\kappa$ could be obtained from the spectral equation
\begin{equation}
\cos \theta - h_V(i\kappa)=0, \qquad \textrm{for} \qquad \theta<\theta_c.
\end{equation}
Nevertheless, instead of considering one value of the mass for each fixed value of the quasi-momentum, it is possible to define $m=\kappa_{min}$ being $i \kappa_{min}$ the wave vector associated to the bound state with minimum energy for the comb, which happens at $\theta=0$. This value of the wave vector can be computed from
\begin{equation}
\cos 0 - h_V(i\kappa_{min})=0, \qquad \textrm{i.e.}\qquad h_V(i\kappa_{min})=1.
\end{equation}
For another potential $V(z)$ that had more than one bound state in the spectrum for a fixed value of $\theta$, $m\geq \kappa_{min}$ where $\kappa_{min}$ will correspond to the lowest energy state among all the states with negative energy of the spectrum.  Moreover,  the contribution of the states with negative energy in \eqref{eqgen5} would contain as many terms as such states in the spectrum, i.e.  it will be proportional to $ \int_{0}^{\theta_c} d\theta \sum_n (m^2-\kappa_{\theta,n}^2)^{3/2} $.

Now that all the negative energy states of the spectrum have been taken into account, the positive energy states must be added. Their wave vectors $k\in \mathbb{R}^+$ can be calculated from \begin{equation}\cos \theta - h_V(k)=0.
\end{equation} One could  perform the summation by using the Cauchy's residue theorem for a complex integral over a contour which enclose all the real positive zeroes of the spectral function. These zeroes characterise both to the states related to the upper bands and to the portion of the first band in which $E>0$. To sum up, the quantum vacuum energy for the PT comb (and similarly for any other comb with bound states in its spectrum) is given by
\begin{eqnarray}\label{E0PT}
&&\hspace{-15pt} E_0^{PT \, comb} =-  \int_{0}^{\theta_c} \frac{d\theta}{\pi} \frac{(m^2-\kappa_\theta^2)^{3/2}}{6\pi} - \int_{0}^{\pi} \frac{d\theta}{\pi} \oint_\Gamma \frac{dk}{2\pi i} \,\, \frac{(m^2+k^2)^{3/2}}{6\pi} \,  \partial_k \log f_\theta(k),
\end{eqnarray}
where $\Gamma$  is the contour represented in figure  \ref{fig:fig1a}.  If there were no states with negative energy in the spectrum,  $m=0$ in the aforementioned contour and in \eqref{E0PT},  where the first integral would disappear.  
\begin{figure}[H]
\centering
\includegraphics[width=0.5\textwidth]{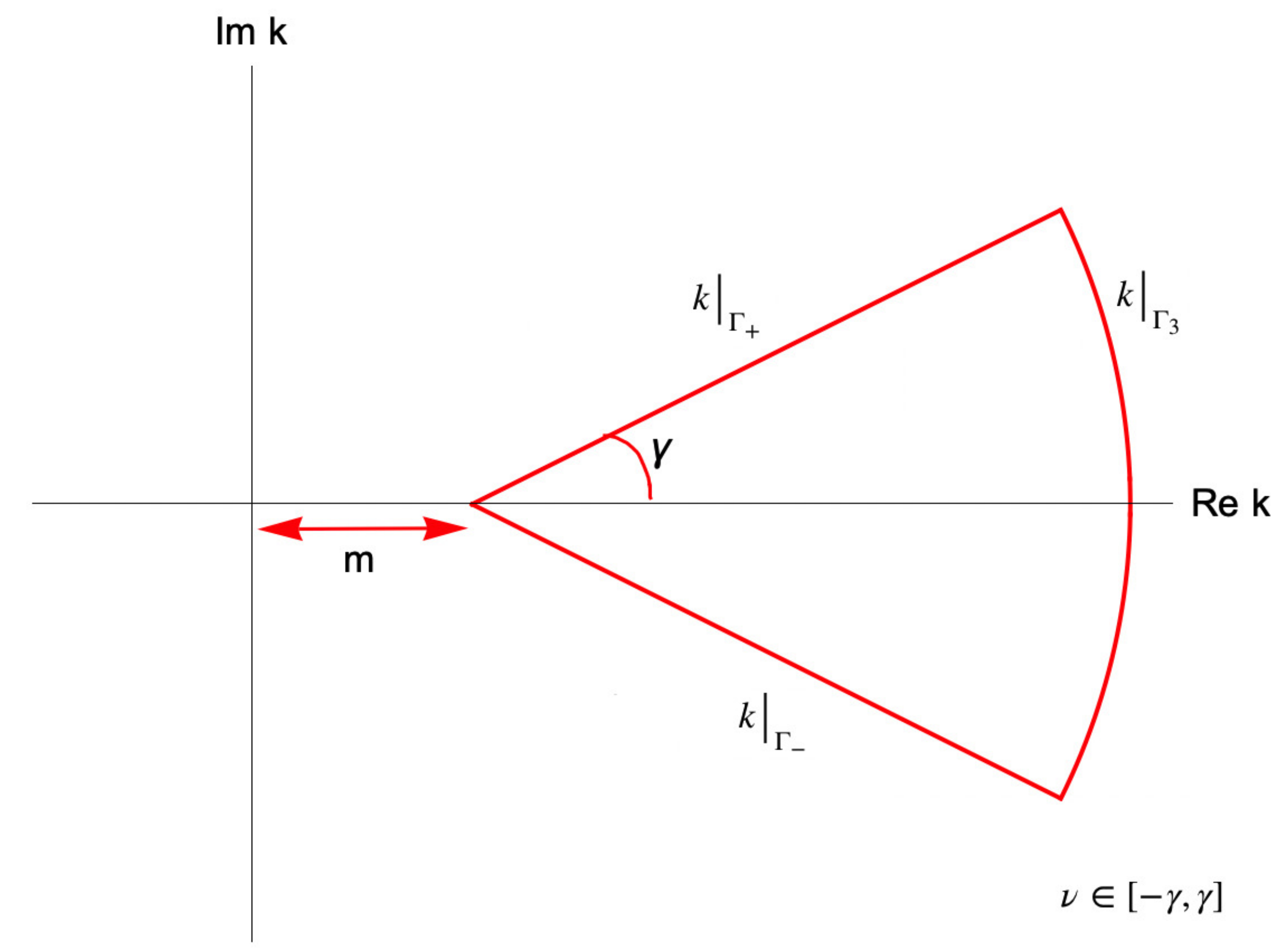}
\caption[\footnotesize Complex contour $\Gamma$ that encloses all the zeroes of $f_\theta(k)$ as $R\to \infty$ when there are bound states in the spectrum]{\footnotesize Complex contour $\Gamma$ that encloses all the zeroes of $f_\theta(k)$ as $R\to \infty$ when there are bound states in the spectrum.  In the contour, $\left. k\right|_{\Gamma_3}=\{(m+R \cos \nu)+ iR \sin \nu, \nu\in[-\gamma, \gamma]\}$ and $\left. k\right|_{\Gamma_\pm}=\{m+\xi e^{\pm i \gamma}, \xi\in [0,R]\}$.  Notice that $R>0$ and $0<\gamma<\pi/2$ are constants.  This contour must be traversed in a counterclockwise sense when integration is carried out.}\label{fig:fig1a}
\end{figure}
Notice that on the one hand,  the secular function $f_\theta(k)$ is a holomorphic function on $k$.  The logarithmic derivative of the secular equation has poles at the zeroes of $f_\theta(k)$, which are the bands in the real axis when summing over the quasi-momentum.  On the other hand, the function $(m^2+k^2)^{3/2}= e^{3 \log_{-\pi}(m^2+k^2)/2}$ is not analytic for $k=|k| e^{i(2p+1)\pi/2}, \, p\in \mathbb{Z}$ providing $|k|\geq m$.  Consequently,  the choice of the contour guarantees that only the real zeros of the spectral function are summed over when performing the integral in $k$ of the above expression and that the branch cuts and bound states in the positive imaginary axis are disregarded. Take into account that the bound states have already been considered in the first summand of \eqref{E0PT}. Furthermore, in the limit $R \to \infty$, the integral over the circumference arc of the contour $\Gamma$  goes to zero.  It can be easily seen when considering a regularisation procedure by means of the heat trace, which is the Mellin transform of the zeta function.  In this way, explained in \cite{JMMCthesis}, one considers
\begin{equation}
\lim_{R\to \infty}\oint_{\Gamma}\frac{dk}{2\pi i} \, e^{- y (k^2+m^2)}\,  (k^2+m^2)^{3/2}\, \partial_k \log f_\theta(k),
\end{equation}
where it is easy to show that the integral over the arc of the contour cancels out thanks to the decaying exponential function. Note that $y$ is an ultraviolet regulator parameter such that the theory is finally recovered by taking the limit $y\to 0$. In conclusion,  integrating over the whole contour is equivalent to integrating over the  straight lines  $\left. k\right|_{\Gamma_\pm}\equiv \xi_\pm=m+\xi e^{\pm i \gamma}$ with $\xi\in[0,R]$. Thus,
\begin{eqnarray}\label{intregular}
E_0^{PT \, comb} =  -\int_{0}^{\theta_c}\! \frac{d\theta}{\pi} \frac{(m^2-\kappa_\theta^2)^{3/2}}{6\pi} - \lim_{R \to \infty} \int_{0}^{\pi} \!\!\frac{d\theta}{\pi}  \!\! \int_{0}^R \!\! \frac{d\xi}{12\pi^2 i } \!\! \!\!\! \! \! && \left[ -(m^2+\xi_+^2)^{3/2} \, \partial_\xi \log f_\theta(\xi_+)  \right. \nonumber\\
 && \left.+(m^2+\xi_-^2)^{3/2} \, \partial_\xi \log f_\theta(\xi_-) \right]
\end{eqnarray}
At this point the integral does not yield a finite result due to the contribution to the dominant and subdominant divergences of the modes in the orthogonal direction.   Divergences in this direction would be different from the ones in the parallel directions,  because now the space is not longer a free one because of the lattice structure.  Hence the method for subtracting divergences now is going to be a little bit different (see \cite{Asorey2013} for a review).  The dominant divergence is the contribution of the energy density of the field theory in the bulk (in this case the space between PT wells,  which is proportional to the infinite length of the y-axis times the lattice spacing $a$), and does not depend on the boundary conditions.  Hence, first of all it is necessary to remove it by subtracting from the integrand a term proportional to the lattice spacing.  In this way,  when integrating $\int_0^\infty d\xi [-(m^2+\xi^2)^{3/2} \, \partial_\xi \log f_\theta(\xi)+ (m^2+\xi^2)^{3/2}  a]$  the contribution of the leading divergence of the vacuum energy induced from fluctuations of the fields in the bulk no longer appears.  But the arising result is still divergent due to the surface density energy associated to the wells with infinite area.  This divergence does not depend on the distance between wells.  To remove it,  what can be done is subtracting the vacuum energy of an
identical system with the same boundary conditions defined over a fixed lattice spacing $a_0$,  which acts as a reference parameter \cite{Asorey2013}.   Notice that $a_0$ is an auxiliary parameter used to define a physical observable that does not diverge in the ultraviolet limit. This observable is the finite component of the vacuum energy density,  which defines the strength of the Casimir effect. Therefore,  when subtracting from this magnitude its value when the plates are at a given distance,  these subdominant divergent terms cancel completely.  In this way,  what remains, i.e.  the result of $$\int_0^\infty d\xi [-(m^2+\xi^2)^{3/2} \, \partial_\xi \log f_\theta(\xi)+ (m^2+\xi^2)^{3/2}  a-(m^2+\xi^2)^{3/2} a_0+(m^2+\xi^2)^{3/2} \partial_\xi \log f_\theta(\xi,a_0)]$$ is free of either dominant and subdominant divergences,  and thus finite.   Furthermore, the finite contribution of the subtraction (those terms depending on $a_0$) cancels out when $a_0$ tends to infinity.  So the final result does not depend on the reference length used to subtract the infinite parts.  To sum up, the finite contribution of the quantum vacuum energy in the problem is the limit $a_0 \to \infty$ of
\begin{eqnarray}\label{limita0toinfty}
\int_{0}^{\pi} \!\!\frac{d\theta}{\pi}  \int_{0}^\infty \!\! \frac{d\xi}{12\pi^2 i }   \left[ (m^2+\xi_+^2)^{3/2}\, \ell_+(\xi)-(m^2+\xi_-^2)^{3/2} \, \ell_-(\xi) \right].
\end{eqnarray}
with $$\ell_\pm(\xi)=a-\partial_\xi \log f_\theta(\xi_\pm, a)-a_0+\partial_\xi \log f_\theta(\xi_\pm, a_0).$$ There is a subtlety here that is important to clarify.  One could think that the result of the aforementioned integral is arbitrary because divergences in QFT can only be subtracted up to a constant when one renormalises the theory.  When renormalising loop diagrams,  infinite contributions are eliminated leaving a finite term which is not uniquely determined.  Instead, each type of divergence in a loop diagram produces a parameter (analogous to the coupling constant) which must be experimentally adjusted.  Notice that here,  I am not renormalising the theory.  I have only regularised the integral \eqref{intregular}.  This means that I have isolated from the total vacuum energy (which is the infinite quantity given by the sum of the quantum vacuum interaction energy or Casimir energy,  the density energy of the theory in the bulk and the self-energy of objects of infinite surfaces) the only contribution that depends on the distance between the Dirac plates or PT wells,  which is now a finite quantity that is actually measured experimentally. That is,  by eliminating the divergences of the previous integral \eqref{intregular} I have not eliminated them from the theory but I have put them into the other divergent terms of the vacuum energy,  which I am not interested in studying at the moment.  If a were to do so, I would have to apply renormalisation procedures.  The important conclusion here is that after zeta regularisation,  the answer to the originally divergent integral \eqref{intregular} is unique,  and it has been shown to be in exact agreement with the experimental results of the Casimir effect (see the comparison between experimental and theoretical predictions after regularisation in \cite{Asorey2013}). 

After performing the limit $a_0 \to \infty$ in \eqref{limita0toinfty} the  quantum vacuum interaction energy for the  P\"oschl-Teller comb yields:
\small
\begin{eqnarray}\label{E0PTcombren}
&&\hspace{-10pt}E_0^{PT \, comb}  = -\int_{0}^{\theta_c} \frac{d\theta}{\pi} \frac{(m^2-\kappa_\theta^2)^{3/2}}{6\pi} - \int_{0}^{\pi} \frac{d\theta}{\pi} \!\int_{0}^\infty \! \frac{d\xi}{12\pi^2 i}\left[\Omega(\xi_+)-\Omega(\xi_-)+\Lambda(\xi_-)\right]
\end{eqnarray}
\normalsize
being 
\small
\begin{eqnarray}
\Omega(\xi_\pm)&=& (m^2+\xi_\pm^2)^{3/2} \, \left(a-\partial_\xi \log [f_\theta(\xi_\pm, a)t(\xi_\pm)]\right),\nonumber\\
\Lambda(\xi_-)&=& (m^2+\xi_-^2)^{3/2} \, i \, 2\,  \partial_\xi \delta(\xi_-).
\end{eqnarray}
\normalsize
Notice that $\delta(\xi)$ is the phase shift obtained from the scattering data by means of the relation $e^{i2\delta(k)}= t^2(k)-r_R(k)r_L(k)$ \cite{Galindobook}.

For the generalised Dirac comb, whenever $w_0>0$ only positive energy bands appear in the sprectrum.  In this case,  the  quantum vacuum interaction energy between parallel Dirac plates can be computed as:
\begin{eqnarray}\label{E0PTcombren}
 \! \! \! \! \! \!  \! \! \! E_0^{\delta\delta' \, comb} \! =- \! \int_{0}^{\pi}  \! \!\frac{d\theta}{\pi} \!\int_{0}^\infty \! \! \! \! \frac{d\xi}{12\pi^2 i}\!\!\!\!&&\left[\tilde{\Omega}(\xi_+)-\tilde{\Omega}(\xi_-) +\, i \, 2\, \xi_-^{3}\,   \partial_\xi \delta(\xi_-)\right]
\end{eqnarray}
being $\delta(\xi)$ the phase shift and 
\begin{equation}
\tilde{\Omega}(\xi_\pm)= \xi_\pm^{3} \, \left(a-\partial_\xi \log f_\theta(\xi_\pm, a)-\partial_\xi \log t(\xi_\pm)\right).
\end{equation}

There is another interesting caveat which becomes apparent when one subtracts either the dominant and subdominant divergences directly in the integrand of the closed complex integral given in equation \eqref{E0PT}. This detail can be also seen from the analogous integrals which appear for combs without negative energy bound states in the spectrum.  When considering this last case, only for simplicity, one has to study
\begin{equation}
\oint_\Gamma \frac{dk}{2\pi i} \,\,  k^3 \,  \left(a-\partial_k \log f_\theta(k) - \partial_k \log t(k)\right),
\end{equation} 
with $\Gamma$ the complex contour given in Figure \ref{fig:fig1a} for $m=0$, and taking for instance $\gamma=\pi/2$ as done in \cite{Asorey2013}. Scattering theory states that the trasmission coefficient can be written in terms of the phase shift as $t(k)=|t(k)| e^{i \delta(k)}$. And the logarithmic integral of $t(k)$ along the complex contour $\Gamma$ is the increase on the argument of $t(k)$ (i.e. the phase shift), by the Argument Principle\footnote{It is possible to rewrite the logarithmic integral of $t$ as $\oint d \log t = \oint d \log |t(k)| +i \oint d\delta(k) $. The differential $d \log |t(k)|$ is exact so its integral over a closed contour is zero. But the differential $d\delta(k)$ is closed but not exact. Its integral is the increase of the phase shift along the contour.} \cite{Gamelinbook}. Consequently, subtracting the subdominant divergence implies the emergence of the phase shift derivative. This would be relevant for obtaining the 3D analogous of the Dashen-Hasslacher-Neveu formula  in one-dimensional spaces \cite{DHN1974a, DHN1974b}, viz.
\begin{eqnarray}
E_0^{DHN}= \frac{1}{2}\int \frac{d\omega}{2\pi} \omega\, \frac{d\delta(\omega)}{d\omega}
\end{eqnarray} 
 in the limit $a \to \infty$,  in which  one considers the contribution to the zero point energy of the states of the spectra in the continuum.

\subsection{Generalised Dirac  comb}
Figure \ref{fig:en1} shows the quantum vacuum interaction energy $E_0^{\delta \delta' \, comb}$ as a function of the distance between nodes $a$ for different values of the $\delta$-$\delta'$ couplings.   In order to analyse a situation different from that of the PT comb,  only the $w_0>0$ case is considered here.  This means that there are no states with negative energy in the spectra.  The $\delta$-$\delta'$ potentials placed at each lattice node can hold one bound state thus mimicking atoms that have lost their most external electron. The classical force between them is repulsive (all the atoms have positive charge).  However, as can be seen in Figure \ref{fig:en1}, the quantum vacuum interaction energy produced by the phonon field can be positive, negative or zero.  
\begin{figure}[h]
\centering
\includegraphics[width=0.45\textwidth]{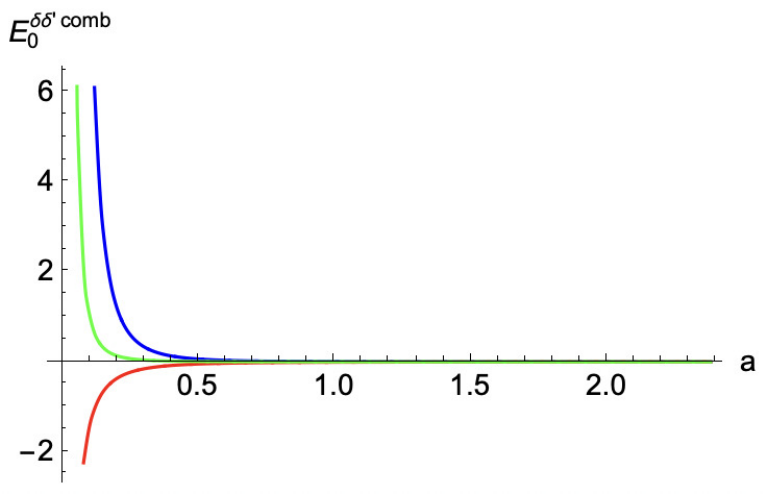}\quad \includegraphics[width=0.45\textwidth]{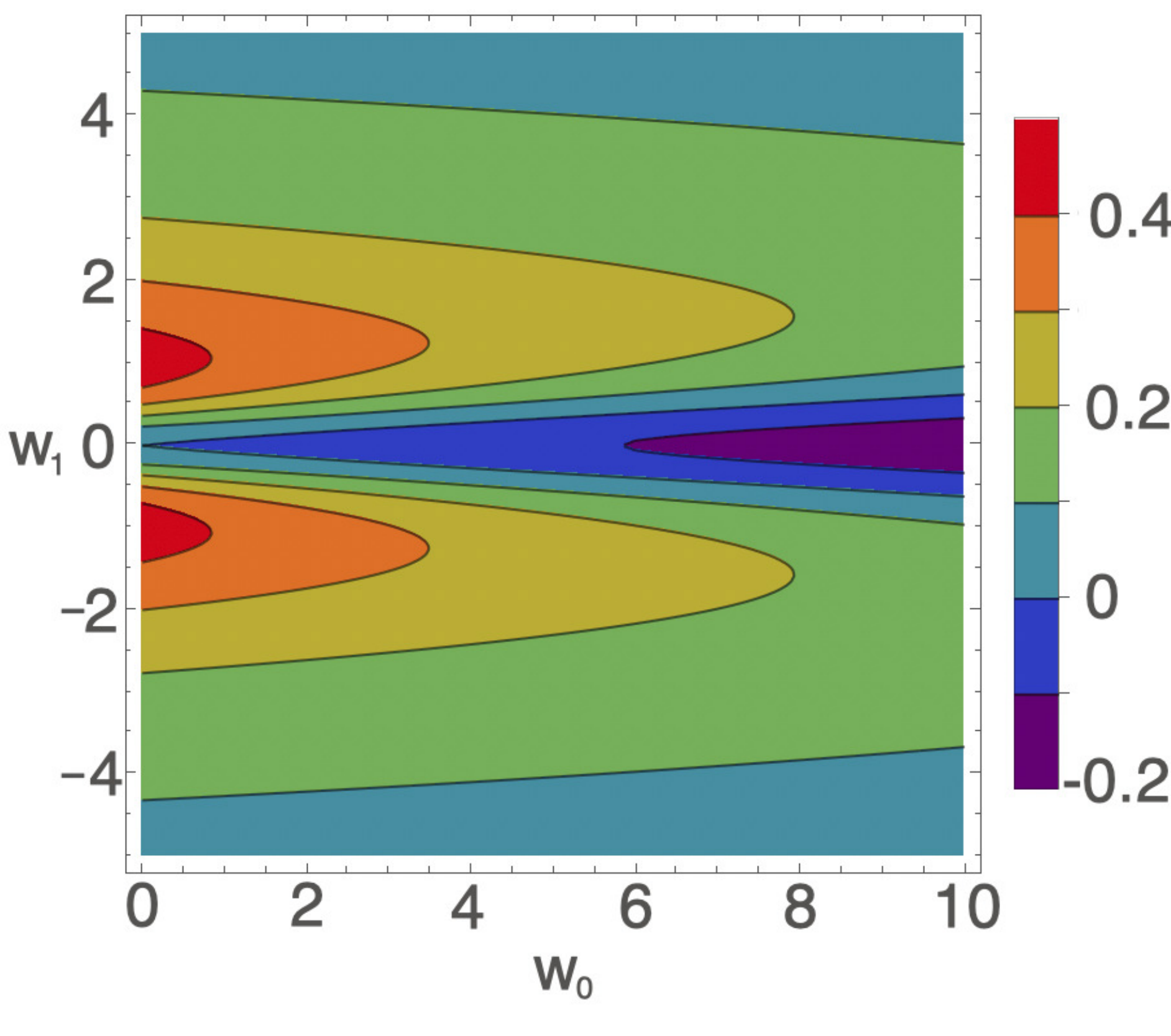}
\caption{\footnotesize Left: Quantum vacuum interaction energy $E_0^{\delta \delta' \, comb}$ as a function of the distance between nodes $a$ for different values of the $\delta$-$\delta'$ couplings: $w_0=2,  w_1=-1$ (blue), $w_0=0,w_1=-7$ (green),  $w_0=10, w_1=0$ (red). Right: Quantum vacuum interaction energy $E_0^{\delta \delta' \, comb}$ for $a=0.3$ in the space of $\delta$-$\delta'$ couplings.}
\label{fig:en1}
\end{figure}

Similarly, the quantum vacuum interaction force between nodes in this three-dimensional system can be repulsive, attractive or zero, because it is obtained from $-\partial E_0^{\delta \delta' comb}/\partial a$. The negative values of the quantum vacuum force implies the reduction of the repulsive classical one, giving rise to a smaller lattice spacing. On the opposite, the positive values mean that the classical repulsion is enhanced and the lattice spacing in the crystal becomes bigger. In all the cases shown in figure  \ref{fig:en1} the quantum vacuum interaction energy becomes zero as $a\to\infty$ because  in the limit case where the lattice nodes are far apart, the phonons propagation initiated at one node ends before reaching the following one and there is no interaction  between nodes.  This is qualitatively in agreement with the results obtained for one-dimensional Dirac chains in \cite{Santamaria2019}. 

\subsection{P\"oschl-Teller comb}
\label{subsection3.2.2}
This second example of crystal is appealing because there is always a band of negative energies regardless of  the value of $a$ and $\epsilon$,  as shown in Figure \ref{fig:spectrumPT}.  Figure \ref{fig:e0PTcomb1} shows $E_0^{PT \, comb}$ for different values of the lattice spacing $a$ and the compact support $\epsilon$ of the P\"oschl-Teller potential which composes the lattice.  Either for the Dirac comb and the P\"oschl-Teller one, the classical force between the lattice nodes is repulsive.  However, it has been found that at zero temperature, the  quantum vacuum interaction energy produced by the phonon field in the PT comb always takes negative values (as can be seen in Figure \ref{fig:e0PTcomb1}). It can be checked that the Casimir pressure between nodes is always positive,  which enhances the repulsive classical force between lattice nodes. 
\begin{figure}[H]
\centering
\includegraphics[width=0.45 \textwidth]{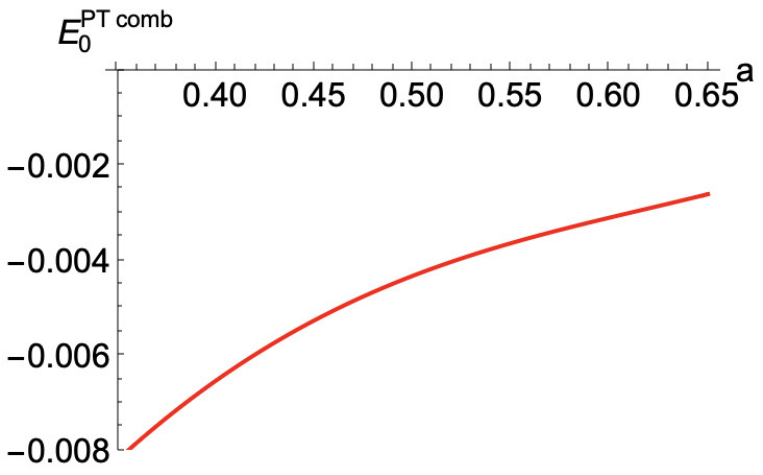}\quad  \includegraphics[width=0.45 \textwidth]{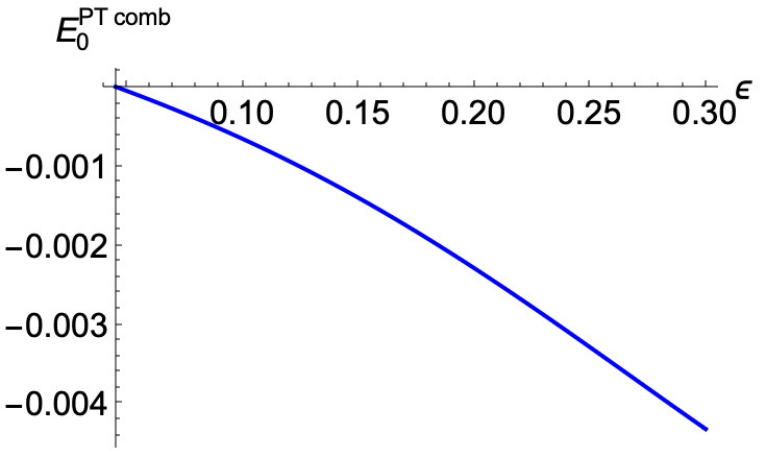}
\caption{\footnotesize Left: Quantum vacuum interaction energy $E_0^{PT\, comb}$ for $\epsilon=0.05$ as a function of the lattice spacing $a$.  Right: Quantum vacuum interaction energy $E_0^{PT\, comb}$ \eqref{E0PTcombren} for the lattice spacing $a=0.5$ as a function of the compact support $\epsilon$.}
\label{fig:e0PTcomb1}
\end{figure}

Note that in order to perform the divergence subtraction explained above in this section,  it must be considered that the ``upside-down tiled roof" built by the PT potential must be much deeper than the separation distance between two parallel P\"oschl-Teller wells. The reason is that this procedure is valid when considering infinite plates (as in the original paper \cite{Asorey2013}) or potentials that may resemble this configuration.  Otherwise,  the effects caused by the plate edges would be appreciable,  and other regularisation methods that go beyond the scope of this work would have to be considered.  For instance,  one could use spectral functions associated to the Laplacian operator \cite{Kirstenbook}.   Another option would be to multiply the potential \eqref{PTpot}  by a positive numerical constant to make the well deeper.  This would allow,  after recalculation of the spectrum of the new comb,  to consider larger values of $a,\epsilon$ while still using the same divergence subtraction method as here.

\section{Free energy and entropy at finite temperature}
\label{sec:temp}
The temperature dependent part of the total Helmholtz free energy  is computed as the summation of the Boltzmann factors 
\begin{equation}
B=T \log (1-e^{-\omega/T})
\end{equation} over the quantum field modes that form the comb spectrum. In order to perform this summation one has to integrate over the parallel momenta and sum over the discrete set of momenta of the orthogonal modes,  as explained in the previous section, viz. :
\small
\begin{eqnarray}
&&\hspace{-10pt}\frac{\bigtriangleup _T\mathcal{F}}{A} =\!\!\!\ \sum_{\omega^2 \in \sigma (\hat{K}_{comb})} \!\!\!\!\!\!\! \!\!B(\omega, T)= \int_{-\pi}^{\pi} \frac{d\theta}{2\pi} \sum_{\omega^2 \in \sigma (\hat{K}_{\theta})} \!\!\!\!\!\! B(\omega, T)= \int_{-\pi}^{\pi}\!\frac{d\theta}{2\pi}  \int_{\mathbb{R}^{2}} \frac{d^{2} \vec{k_\|}}{{\color{black}(2\pi)^2}}  \sum_{k\in Z(f_\theta) }\!\!\!\! B\left(\sqrt{k^2+\vec{k_\parallel}^2+m^2}, T\right).\nonumber \\&&
\end{eqnarray}
\normalsize
Whenever there are sates with negative energy in the quantum mechanical problem, this last summation splits into the part of the band with states of negative energies and the bands with positive energies.  For instance, in the PT comb problem one can write 
\begin{eqnarray}
&& \frac{\bigtriangleup _T\mathcal{F}^{PT comb}}{A} =\int_{0}^{\theta_c}\frac{d\theta}{\pi} \!\!\int_{\mathbb{R}^2} \frac{d\vec{k}_\parallel^2}{(2\pi)^2} B\left(\sqrt{m^2-\kappa^2+\vec{k}_\parallel^2}, T\right) \nonumber\\
&+& \int_{0}^{\pi}\frac{d\theta}{\pi} \int_{\mathbb{R}^2} \frac{d\vec{k}_\parallel^2}{(2\pi)^2}  \sum_{k\in Z(f_\theta)} \!\! B\left(\sqrt{m^2+k^2+\vec{k}_\parallel^2}, T\right)
\end{eqnarray}
\normalsize
being  $\omega=\sqrt{k^2+\vec{k_\parallel}^2+m^2}$ the energy of the one-particle states of the Quantum Field Theory.

The integration over the parallel momenta can be commuted with the summation over the discrete transverse momenta in order to perform the integral first by using {\it Mathematica} and obtain
\begin{eqnarray}
&& I_3(\sqrt{m^2+k^2}, T) :=T\int_{\mathbb{R}^{2}} \frac{d^{2} \vec{k_\|}}{{\color{black}(2\pi)^2}}\,   \log \left( 1-e^{\textstyle{-\frac{\sqrt{k_\|^2+k^2+m^2}}{ T}}}\right)\nonumber\\
&&=\frac{T }{2\pi} \int_0^\infty dk_\| \, k_{\|} \,   \log \left( 1-e^{-\lambda \sqrt{\frac{k_\|^2}{k^2+m^2}+1}}\right)= -  \frac{T^3}{2\pi} \left[\lambda \,\, \text{Li}_2\left(e^{-\lambda}\right)+\! \text{Li}_3\left(e^{-\lambda}\right)\right]\label{i3-def}
\end{eqnarray}
being $\lambda=\sqrt{k^2+m^2} /T$ and $\text{Li}_s(\mathsf{z})$  the polylogarithmic function of order $s$ \cite{NISTbook}.  Consequently,
\begin{eqnarray}\label{DTligcomb}
&&\frac{\bigtriangleup _T\mathcal{F}^{PT comb}}{A} = \!\! \int_{0}^{\theta_c}\frac{d\theta}{\pi}  I_3\left(\sqrt{m^2-\kappa_{\theta}^2}\,\, , T\right)+\int_{0}^{\pi}\frac{d\theta}{\pi} \sum_{k\in Z(f_\theta) }   I_3\left(\sqrt{m^2+k^2}, T\right)  \nonumber\\
&&=  \int_{0}^{\theta_c}\frac{d\theta}{\pi}  I_3\left(\sqrt{m^2-\kappa_{\theta}^2}\,\, , T\right)+\lim_{R \to \infty} \int_{0}^{\pi}\frac{d\theta}{\pi} \oint_\Gamma  \frac{dk}{2\pi i} I_3\left(\sqrt{m^2+k^2}, T\right) \partial_k \log f_\theta(k).
\end{eqnarray}
\normalsize
Notice that in the last step, the summation over the discrete set of orthogonal momenta can be computed as a complex integral over  the contour given in Figure \ref{fig:fig1a} by the Cauchy's theorem.  

For combs whose spectrum do not present negative energy bands,  for example the generalised Dirac comb,  the temperature dependent part of the Helmholtz free energy $\bigtriangleup _T\mathcal{F}^{\delta\delta' comb}/A$ can be computed as
\begin{eqnarray}\label{DTcomb}
\!\!  \lim_{R \to \infty} \int_{0}^{\pi}\frac{d\theta}{\pi}  \oint_\Gamma \frac{dk}{2\pi i} \, I_3(k, T)  \,  \partial_{k} \log f_\theta(k),
\end{eqnarray}
 where $\Gamma$ is the contour in Figure \ref{fig:fig1a} for $m=0$. The result of the integration in \eqref{DTcomb} and \eqref{DTligcomb} does not depend on the angle $\gamma$ taken in the contour, according to the residue theorem. The major advantage of choosing the angle of the contour such that $0<\gamma<\pi/4$, is that it avoids the oscillations of the integrand caused by the zeroes of the secular function on the real axis.  Moreover, the integrand has an exponential decrease which makes numerical evaluation easier. Nonetheless, it is possible to choose either $\gamma=0$ recovering the well-known real frequencies approach, or $\gamma=\pi/2$ to work on the Matsubara representation. 

The combination of polylogarithms which defines $I_3$ is analytic over the entire complex plane,  and the non-analytic points of the argument of the polylogarithm are outside the contour,  as explained in the previous section.  Moreover, the logarithmic derivative of the secular function has poles at the zeroes of $f_\theta(k)$, which are the bands in the real axis when summing over the quasi-momentum. Furthermore, in the limit $R \to \infty$, the integral over the circumference arc of the contour $\Gamma$  goes to zero following the reasoning given in the previous section and taking into account that
\begin{eqnarray}
&&\lim_{R\to \infty}I_3(\sqrt{m^2+R^2 e^{i 2\nu}}, T)=\lim_{R\to \infty} \frac{T^3 \left[-1-R e^{i \nu} /T\right]\, e^{-R e^{i \nu}/T}}{2\pi} =0.
\end{eqnarray}

In conclusion, integrating over the whole contour is again equivalent to integrating over the two straight lines   $\xi_\pm=\xi e^{\pm i \gamma} +m$ with $\xi\in[0,\infty)$, $m>0$ and $\gamma \in (0, \pi/2)$. So finally
\begin{eqnarray}
&&\hspace{-10pt}\frac{\bigtriangleup _T\mathcal{F}^{PT comb}}{A} =\int_{0}^{\theta_c}\frac{d\theta}{\pi}  I_3\left(\sqrt{m^2-\kappa_{\theta}^2}\, \, , T\right)\nonumber\\
&&\hspace{-10pt}+\int_{0}^{\pi}\frac{d\theta}{\pi}\!\!    \int_{0}^{\infty} \!\! \frac{d\xi}{\pi} \textrm{Re} \, \left[i I_3\left(\sqrt{m^2+\xi_+^2}, T\right) \partial_\xi \log f_\theta(\xi_+) \right] 
\end{eqnarray}
for combs with negative energy bands,  and 
\begin{eqnarray}
\!\!\frac{\bigtriangleup _T\mathcal{F}^{\delta \delta' comb}}{A} \!\! &=&\!\! \int_{0}^{\pi}\!\!\frac{d\theta}{\pi} \!\!  \int_{0}^{\infty}\!\!\frac{d\xi}{\pi} \textrm{Re}  \!\left[i I_3\!\left(\xi_+, T\right) \partial_\xi \log f_\theta(\xi_+) \right] 
\end{eqnarray}
for combs with only positive energy bands.  These integrals can be  computed numerically by using \textit{Mathematica} for any finite temperature $T$. The numerical results for both the Dirac and P\"oschl-Teller comb in three dimensions are given in figure \ref{fig:Fcomb3D}. It can be easily checked that $\bigtriangleup _T\mathcal{F}/A$ takes always negative values which decrease rapidly with increasing temperature.  
\begin{figure}[h]
\centering
\includegraphics[width=0.45\textwidth]{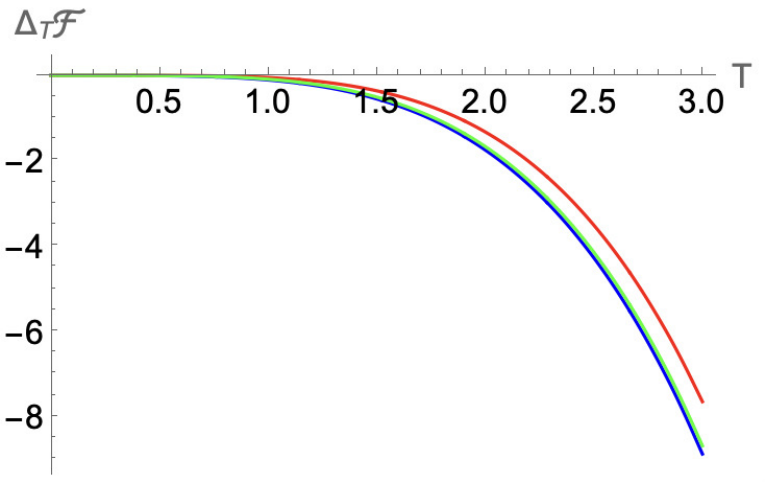}\quad \includegraphics[width=0.45\textwidth]{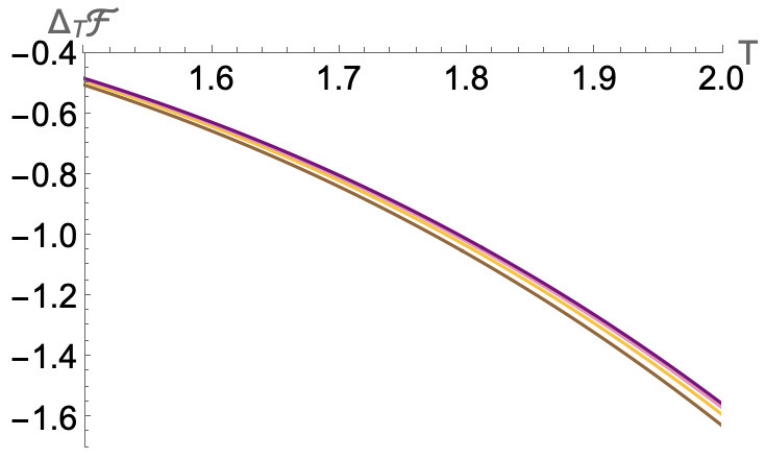}
\caption{\footnotesize  $\bigtriangleup _T\mathcal{F}/A$ for 3D combs (Dirac comb on the plot on the left and PT comb on the right) as a function of the temperature for $a=1$.  Left: $w_0=0.1, w_1=-5$ (blue),  $w_0=8, w_1=0$ (red), $w_0=3, w_1=-2$ (green).  Right: $\epsilon=0.25$ (brown),$ \epsilon=0.5$ (yellow),$\epsilon=0.75$ (pink),$\epsilon=0.98$ (purple).}
\label{fig:Fcomb3D}
\end{figure}

From the total Helmholtz free energy  one could obtain the one-loop quantum corrections of the entropy and the Casimir pressure by means of
\begin{equation}
\frac{S}{A} = -\frac{d}{dT} \left( \frac{\bigtriangleup_T \mathcal{F}}{A}\right),  \qquad P= -\frac{d}{d\, a}\left( \frac{\bigtriangleup_T \mathcal{F}}{A}\right).
\end{equation}
The results are shown in Figures \ref{fig:Scomb3D} and \ref{fig:Pcomb3D}. As can be seen, the entropy is a monotonically increasing function of the temperature, meaning that the quantum systems are thermodynamically stable. Negative corrections to the entropy are related to instabilities in the system \cite{Thirring1970}.  Finally, the Casimir force per unit area takes positive values too, which can be interpreted as the lattice spacing increases and the unit cell blows up as a consequence of the quantum interaction at non trivial temperatures.

\begin{figure}[h]
\centering
\includegraphics[width=0.45\textwidth]{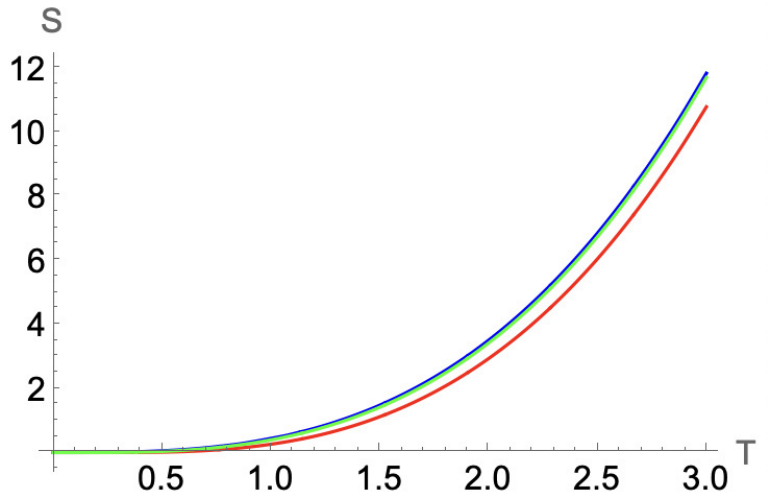}\qquad \includegraphics[width=0.45\textwidth]{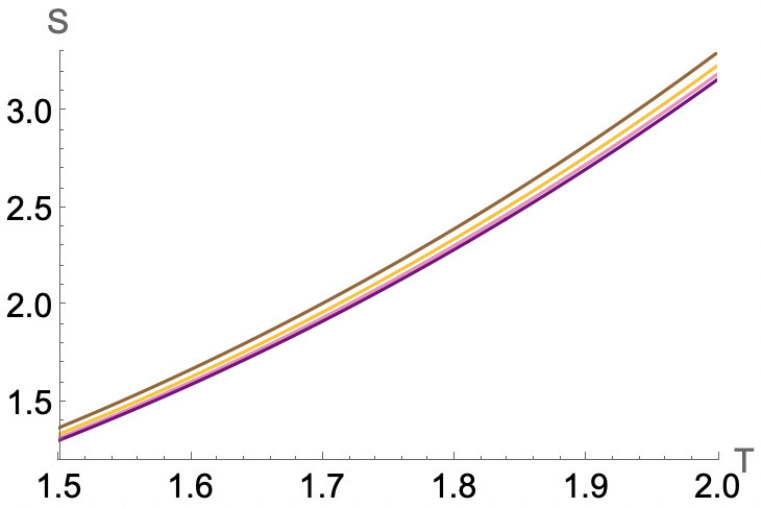}
\caption{\footnotesize  Entropy per unit area for 3D combs (Dirac comb on the plot on the left and PT comb on the right) as a function of the temperature for $a=1$.  Top: $w_0=0.1, w_1=-5$ (blue),  $w_0=8, w_1=0$ (red), $w_0=3, w_1=-2$ (green). Bottom: $ \epsilon=0.25$ (brown), $\epsilon=0.5$ (yellow),$ \epsilon=0.75$ (pink),$\epsilon=0.98$ (purple).}
\label{fig:Scomb3D}
\end{figure} 

\begin{figure}[H]
\centering
\includegraphics[width=0.45\textwidth]{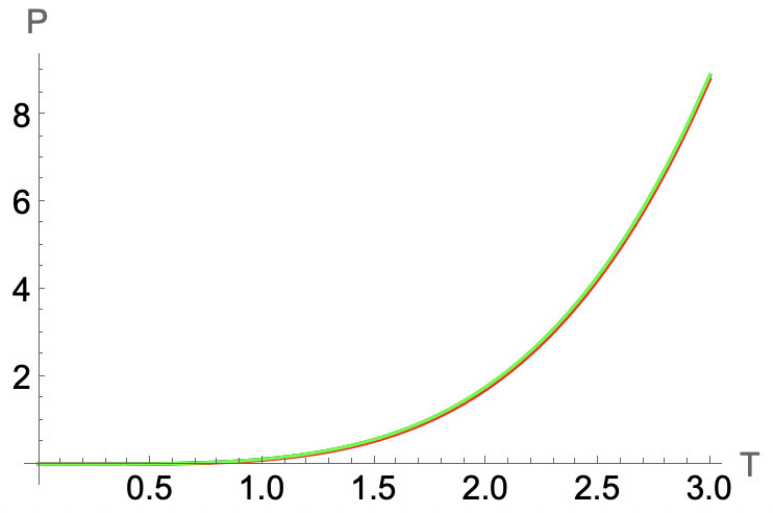}\qquad \includegraphics[width=0.45\textwidth]{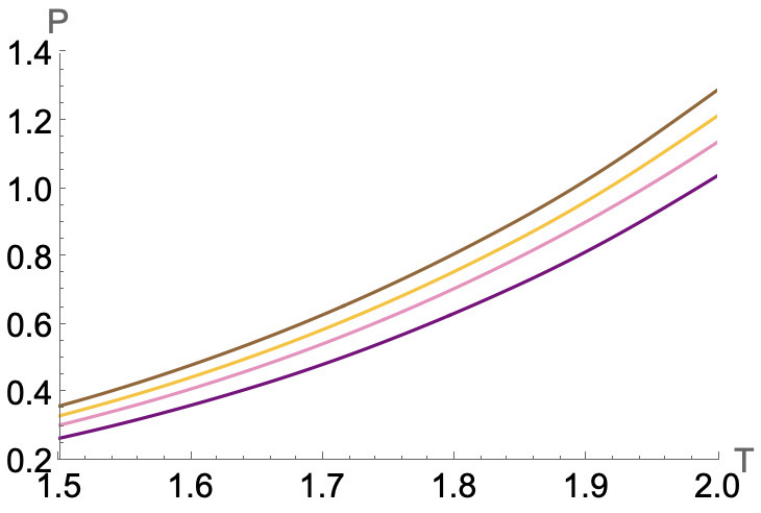}
\caption{\footnotesize Casimir pressure for 3D combs (Dirac comb on the plot on the left and PT comb on the right) as a function of the temperature for $a=1$.  Left: $w_0=0.1, w_1=-5$ (blue),  $w_0=8, w_1=0$ (red), $w_0=3, w_1=-2$ (green). Right: $ \epsilon=0.25$ (brown),$\epsilon=0.5$ (yellow),$ \epsilon=0.75$ (pink),$\epsilon=0.98$ (purple).}
\label{fig:Pcomb3D}
\end{figure} 

Notice that if the plates had a finite area, instead of performing an integration over the whole $\mathbb{R}^2$, one would make a summation over a discrete set of parallel momenta.  For instance, in the generalised Dirac comb,  if Dirichlet boundary conditions were imposed over the edges of a squared $\delta \delta'$ plate of length $b$, the plates become completely opaque.  Consequently,  the field must fulfil the relations 
\begin{equation}
\phi(x=\pm b)=\phi(y=\pm b)=0,
\end{equation}
 and the summation of the Boltzmann factors over $\omega^2 \in \sigma(\hat{K}_{comb})$ would be

\begin{equation}
 \int_0^\pi\!\!  \frac{d\theta}{\pi} \sum_{n_x,n_y=1}^\infty \sum_{k\in Z(f_\theta) }\!\!  \!\! B\left(\! \sqrt{k^2+\frac{n_x^2 \pi^2}{b^2}+ \frac{n_y^2 \pi^2}{b^2}}, T\right). 
\end{equation}
 This type of cases are left for future further investigation.

\section{Conclusions}
The spectrum of allowed energy bands and forbidden gaps for chains embedded in a 3+1 dimensional spacetime has been studied. The periodic background is built as the repetition in one of the three spatial dimensions of individual potentials placed at the lattices nodes and with compact support smaller than the lattice spacing.  The analysis has been particularised for the  P\"oschl-Teller comb and the generalised Dirac comb.  To this effect,  the spectrum of modes for the phonons in the primitive cell interacting with the individual potential centred at the middle of this interval has been solved.  The spectrum has been compared with that of the one-dimensional pure Dirac delta chain \cite{Santamaria2020b}.   While the values of the coefficients of the Dirac delta potential and its first derivative establish whether there are or not negative energy bands in the spectrum of the generalised Dirac comb,   the P\"oschl-Teller comb  always has a negative energy band for certain values of the quasi-momentum of the first Brillouin zone,  independently of either the magnitude of the compact support of the individual potential and the value of the lattice spacing.  In both types of lattices there are always bands of positive energies.  Determining whether or not there are states with negative energy is fundamental when promoting the theory from non-relativistic quantum mechanics to Quantum Field Theory.  For those lattices with negative energy bands,  a mass term must be introduced to ensure unitarity in the associated QFT.  The values of such a mass of the fluctuations have been determined with complete generality from the band spectra.

The quantum vacuum interaction energy $E_0$ between nodes of the comb has been computed by using a method  based on the heat equation kernel  regularisation.  The quantum vacuum energy at zero temperature thus calculated is the one-loop quantum correction to the classical repulsive elastic forces produced by the quantum scalar field of the phonons.  In the case of the generalised Dirac comb at zero temperature $E_0$, and thus the Casimir pressure between nodes,  takes positive,  negative,  and zero values depending on the parameters that characterise the lattice.  Consequently, the quantum vacuum force can be repulsive, attractive,  or zero. This means that the lattice spacing can be increased, decreased,  or remain unchanged with respect to its classical analogue as a result of this quantum interaction.  In contrast,  the Casimir pressure between nodes is always positive for the P\"oschl-Teller comb at zero temperature,  so that the classical repulsive force between lattice nodes is enhanced.

The thermal correction to the quantum vacuum energy, the entropy and the Casimir pressure at finite non-zero temperature has been derived in a convergent representation based in complex Cauchy integrals,  which major advantage is that turning the integration contour towards the imaginary axis by a finite angle in the complex plane of frequencies avoids large oscillations of the Boltzmann factor.  Furthermore, this method is model-independent in the sense that the choice of the individual potential forming the lattice does not matter as long as its support be smaller than the lattice spacing. Positive one-loop quantum corrections to the entropy appear for both the generalised Dirac comb and the P\"oschl-Teller  one at finite non-zero temperature. This fact means that the classical analogue system is more stable than the quantum one.  On the other hand, the Casimir force between the lattice nodes is always repulsive for both chains when non-trivial temperatures are considered, implying that the primitive cell increases its size due to the quantum interaction of the phonon field at non-trivial temperatures.  

Looking ahead,  there are many interesting questions to explore in relation to the study of quantum scalar fields propagating along lattices.  For instance, the computation of the Green's function is left for further investigations.  From it one can obtain the vacuum expectation value of the $00$-component of the energy momentum tensor,  which provides much more information concerning the energy density distribution than just computing the total quantum vacuum interaction energy.  It would be also an interesting new line of research to consider some type of smooth interactions between phonons and electrons in the lattices since the realistic material are made of electrons.  In this case the spin statistics properties must be taken into account.  In \cite{Cervero2005} (and first two references therein),  one could find a problem similar to the one addressed here but for fermions. In these works, the authors studied an electron in a periodic one-dimensional chain of atoms modelled by an infinite periodic potential of Dirac $\delta$ with different coefficients inside the unit cell. That model was used by Cerver\'o \textit{et al.} to study quantum wire band structures and Anderson localisation.  What is interesting about this example is that,  although there are some deterministic properties due to the presence of electrons that could not be studied when considering phonons, the methodology used by the authors to characterise the spectra from a secular function is completely analogous to the one that is presented here.  This suggests that it may be possible to combine the results presented in this work with those for fermions from Cerver\'o \textit{et al.}  to study a combined electron-phonon model in the future.  Moreover, the study of the one-loop quantum corrections to the frequency of the phonons propagating along lattices can be used to discuss the stability of some hypothetical solutions regarding the vacuum state in quantum chromodynamics.  Since this state is a non-interacting instanton's gas in the dilute limit (also known as Bogomol'nyi-Prasad-Sommerfield or BPS limit) \cite{PhysRevD.17.2717, Mantonbook},  in which there are kinks with different centres of mass separated by a large distance in such a way that there is a negligible coupling between them,  whenever theories which describe in a sufficiently rigorous way the coupling between the axion fluctuations and the QCD vacuum state can be found,  the methods provided here could be useful. This is left for further investigation. 

\medskip

\noindent {\bf Acknowledgements:}This research was supported by the Spanish Ministry of Science and Innovation MICIN with funding from European Union NextGenerationEU (PRTRC17.I1) and by the Regional Government of Castilla y Le\'on (Junta de Castilla y Le\'on,  Spain) through QCAYLE project.  I thank  L.M. Nieto for  interesting discussions on this subject.  I acknowledge the technical computing support received by J.J.  Relancio.

\bibliographystyle{unsrt}
\bibliography{bibliography.bib}

\begin{thebibliography}{10}

\bibitem{Mostepanenkobook}
A.~A. Grib, S.~G. Mamayev, and V.~M. Mostepanenko.
\newblock {\em Vacuum quantum effects in strong fields}.
\newblock Friedmann Laboratory Publishing, St. Petesburg, Russia, 1994.

\bibitem{Miltonbook}
K.~A. Milton.
\newblock {\em Physical Manifestations of Zero-Point Energy. The Casimir
  Effect}.
\newblock World Scientific, Singapore, Singapore, 2001.

\bibitem{Bordagbook}
M.~Bordag, G.~L. Klimchitskaya, U.~Mohideen, and V.~M. Mostepanenko.
\newblock {\em Advances in the Casimir effect}.
\newblock Oxford Sciences publications, Oxford, U.K., 2009.

\bibitem{Casimir1948}
H.~B.~G. Casimir.
\newblock On the attraction between two perfectly conducting plates.
\newblock {\em Proc. K. Ned. Akad. Wet.}, 51(793), 1948.

\bibitem{Spaarnay1957}
M.~J. Spaarnay.
\newblock {Attractive Forces between Flat Plates}.
\newblock {\em Nature}, 180(334), 1957.

\bibitem{KK2008}
O.~Kenneth and I.~Klich.
\newblock {Casimir forces in a T-operator approach}.
\newblock {\em Phys.Rev. B}, 78:014103, 2008.

\bibitem{Kardar2008}
T.~Emig, N.~Graham, R.~L. Jaffe, and M.~Kardar.
\newblock Casimir forces between compact objects: The scalar case.
\newblock {\em Phys. Rev. D}, 77:025005, 2008.

\bibitem{Alvarez2015}
M.~Asorey, D.~Garc\'ia-\'Alvarez, and J.~M. Munoz-Castaneda.
\newblock Boundary effects in bosonic and fermionic field theories.
\newblock {\em Int. J. Geom. Methods Mod. Phys.}, 12(6):1560004, 2015.

\bibitem{Asorey_2006}
M.~Asorey, D.~Garc\'ia-\'Alvarez, and J.~M. Munoz-Castaneda.
\newblock Casimir effect and global theory of boundary conditions.
\newblock {\em J. Phys. A Math. Gen.}, 39:6127, 2006.

\bibitem{Asorey2013}
M.~Asorey and J.~M. Munoz-Castaneda.
\newblock {Attractive and repulsive Casimir vacuum energy with general boundary
  conditions}.
\newblock {\em Nucl. Phys. B}, 874(3):852--876, 2013.

\bibitem{Santamaria2019}
M.~Bordag, J.~M. Munoz-Castaneda, and L.~Santamar\'ia-Sanz.
\newblock {Vacuum Energy for Generalized Dirac Combs at T = 0}.
\newblock {\em Front. Phys.}, 7, 2019.

\bibitem{Santamaria2020}
M.~Bordag, J.~M. Munoz-Castaneda, and L.~Santamar{\'\i}a-Sanz.
\newblock Free energy and entropy for finite temperature quantum field theory
  under the influence of periodic backgrounds.
\newblock {\em Eur. Phys. J. C}, 80(3):221, 2020.

\bibitem{Santamaria2020b}
M.~Gadella, J.~M.~Mateos Guilarte, J.~M. Munoz-Castaneda, L.~M. Nieto, and
  L.~Santamar{\'\i}a-Sanz.
\newblock Band spectra of periodic hybrid $\delta-\delta^\prime$ structures.
\newblock {\em Eur. Phys. J. Plus}, 135(10):786, 2020.

\bibitem{Ines2016}
K.~V. Shajesh, I.~Brevik, I.~Cavero-Pel\'aez, and P.~Parashar.
\newblock Casimir energies of self-similar plate configurations.
\newblock {\em Phys. Rev. D}, 94(6):065003, 2016.

\bibitem{Donaire2020}
J.~M. Munoz-Castaneda, L.~Santamar\'ia-Sanz, M.~Donaire, and M.~Tello-Fraile.
\newblock {Thermal Casimir Effect with general boundary conditions}.
\newblock {\em Eur. Phys. J. C}, 80:793, 2020.

\bibitem{Kronig1931}
R.~de~L.~Kronig and W.~G. Penney.
\newblock {Quantum Mechanics of Electrons in Crystal Lattices}.
\newblock {\em Proc. R. Soc. Lond. Ser. A}, 130:449, 1931.

\bibitem{Cervero2002}
J.~M. Cerver\'o and A.~Rodr\'iguez.
\newblock {Infinite chain of different deltas: A simple model for a quantum
  wire}.
\newblock {\em Eur. Phys. J. B}, 30:239, 2002.

\bibitem{BordagJPA2020}
M.~Bordag.
\newblock {Conditions for Bose-Einstein condensation in periodic background}.
\newblock {\em J. Phys. A: Math. Theor.}, 53:015003, 2020.

\bibitem{Halevi1999}
I.~Alvarado-Rodr\'iguez, P.~Halevi, and J.~J. S\'anchez-Mondrag\'on.
\newblock {Density of states for a dielectric superlattice: TE polarization}.
\newblock {\em Phys. Rev. E}, 59:3624, 1999.

\bibitem{Hennig1992}
M.~Bordag, D.~Hennig, and D.~Robaschik.
\newblock Vacuum energy in quantum field theory with external potentials
  concentrated on planes.
\newblock {\em J. Phys. A}, 25:4483, 1992.

\bibitem{Fosco2009}
C.~D. Fosco, F.~C. Lombardo, and F.~D. Mazzitelli.
\newblock {Derivative expansion for the boundary interaction terms in the
  Casimir effect: Generalized $\ensuremath{\delta}$ potentials}.
\newblock {\em Phys. Rev. D}, 80:085004, 2009.

\bibitem{Barton2004}
G.~Barton.
\newblock Casimir energies of spherical plasma shells.
\newblock {\em J. Phys. A}, 37:1011, 2004.

\bibitem{Parashar2012}
P.~Parashar, K.~A. Milton, K.~V. Shajesh, and M.~Schaden.
\newblock {Electromagnetic semitransparent $\ensuremath{\delta}$-function
  plate: Casimir interaction energy between parallel infinitesimally thin
  plates}.
\newblock {\em Phys. Rev. D}, 86:085021, 2012.

\bibitem{Gadella2009}
M.~Gadella, J.~Negro, and L.~M. Nieto.
\newblock {Bound states and scattering coefficients of the $-a \delta(x)+b
  \delta^\prime (x)$ potential}.
\newblock {\em Phys. Lett. A}, 373(15):1310, 2009.

\bibitem{Bordagprima2014}
M.~Bordag.
\newblock Monoatomically thin polarizable sheets.
\newblock {\em Phys. Rev. D}, 89:125015, 2014.

\bibitem{Alvarez2013}
J.~J. \'Alvarez, M.~Gadella, L.P. Lara, and F.~H. Maldonado-Villamizar.
\newblock {Unstable quantum oscillator with point interactions: Maverick
  resonances, antibound states and other surprises}.
\newblock {\em Phys. Lett. A}, 377:2510--2519, 2013.

\bibitem{Albeverio1984}
S.~Albeverio, F.~Gesztesy, R.~Hoegh-Krohn, and W.~Kirsch.
\newblock On point interactions in one dimension.
\newblock {\em Journal of Operator Theory}, 12(1):101--126, 1984.

\bibitem{Albeverio2013}
Sergio Albeverio, Silvestro Fassari, and Fabio Rinaldi.
\newblock {A remarkable spectral feature of the Schrödinger Hamiltonian of the
  harmonic oscillator perturbed by an attractive $\delta'$-interaction centred
  at the origin: double degeneracy and level crossing}.
\newblock {\em Journal of Physics A: Mathematical and Theoretical},
  46(38):385305, 2013.

\bibitem{Guilarte2011}
J.~M. Guilarte and J.~M. Munoz-Castaneda.
\newblock {Double-Delta Potentials: One Dimensional Scattering. The Casimir
  Effect and Kink Fluctuations}.
\newblock {\em Int. J. Theor. Phys.}, 50(7):2227--2241, 2011.

\bibitem{PTeller1933}
G.~P\"oschl and E.~Teller.
\newblock {Bemerkungen zur Quantenmechanik des anharmonischen Oszillators}.
\newblock {\em Z. Phys. A}, 83:143–151, 1933.

\bibitem{Negro2016}
D.~\c{C}evik, M.~Gadella, S.~Kuru, and J.~Negro.
\newblock {Resonances and antibound states for the P\"oschl-Teller potential:
  Ladder operators and SUSY partners}.
\newblock {\em Phys. Lett. A}, 380:1600–1609, 2016.

\bibitem{Park2015}
C.~S. Park.
\newblock {Two-dimensional transmission through modified P\"oschl-Teller
  potential in bilayer graphene}.
\newblock {\em Phys.Rev.B}, 92:165422, 2015.

\bibitem{Hartmann2017}
R.~R. Hartmann and M.~E. Portnoi.
\newblock {Two-dimensional Dirac particles in a P\"oschl-Teller waveguide}.
\newblock {\em Sci. Rep.}, 7:11599, 2017.

\bibitem{Tomak2005}
H.~Yildirim and M.~Tomak.
\newblock {Nonlinear optical properties of a P\"oschl-Teller quantum well}.
\newblock {\em Phys. Rev. B}, 72:115340, 2005.

\bibitem{Kroemer1963}
H.~Kroemer.
\newblock A proposed class of hetero-junction injection lasers.
\newblock {\em Proceedings of the IEEE}, 51(12):1782--1783, 1963.

\bibitem{Alferov1963}
Zh.~I. Alferov and R.F. Kazarinov.
\newblock {I. Alferov et al. Authors Certificate 28448, USSR (1963)}.
\newblock {\em Sov. Phys. Solid State}, 9:208, 1967.

\bibitem{Radovanovic2000}
J.~Radovanović, V.~Milanović, Z.~Ikonić, and D.~Indjin.
\newblock {Intersubband absorption in P\"oschl-Teller-like semiconductor
  quantum wells}.
\newblock {\em Physics Letters A}, 269(2):179--185, 2000.

\bibitem{Simons}
B.~Simons.
\newblock {Quantum Condensed Matter Field Theory. Lectures notes. Chapter 1:
  Collective excitations. From particles to fields}, 2017.
\newblock \url{http://www.tcm.phy.cam.ac.uk/~bds10/dir/teaching.html}.

\bibitem{PhysRev.82.664}
J.~Schwinger.
\newblock On gauge invariance and vacuum polarization.
\newblock {\em Phys. Rev.}, 82:664--679, 1951.

\bibitem{ZeljkovicNature2015}
I.~Zeljkovic, Y.~Okada, M.~Serbyn, R.~Sankar, D.~Walkup, W.~Zhou, J.~Liu,
  G.~Chang, Y.J. Wang, M.~Z. Hasan, F.~Chou, H.~Lin, A.~Bansil, L.~Fu, and
  V.~Madhavan.
\newblock {Dirac mass generation from crystal symmetry breaking on the surfaces
  of topological crystalline insulators}.
\newblock {\em Nature Materials}, 14(3):318--324, 2015.

\bibitem{OkadaScience2013}
Y.~Okada, M.~Serbyn, H.~Lin, D.~Walkup, W.~Zhou, C.~Dhital, M.~Neupane, S.~Xu,
  Y.~J. Wang, R.~Sankar, F.~Chou, A.~Bansil, M.~Z. Hasan, S.~D. Wilson, L.~Fu,
  and V.~Madhavan.
\newblock {Observation of Dirac Node Formation and Mass Acquisition in a
  Topological Crystalline Insulator}.
\newblock {\em Science}, 341(6153):1496--1499, 2013.

\bibitem{WangNature2016}
E.~Wang, X.~Lu, S.~Ding, W.~Yao, M.~Yan, G.~Wan, K.~Deng, S.~Wang, G.~Chen,
  L.~Ma, J.~Jung, A.~V. Fedorov, Y.~Zhang, G.~Zhang, and S.~Zhou.
\newblock {Gaps induced by inversion symmetry breaking and second-generation
  Dirac cones in graphene/hexagonal boron nitride}.
\newblock {\em Nature Physics}, 12(12):1111--1115, 2016.

\bibitem{Baggioli_2020}
M.~Baggioli and A.~Zaccone.
\newblock Low-energy optical phonons induce glassy-like vibrational and thermal
  anomalies in ordered crystals.
\newblock {\em Journal of Physics: Materials}, 3(1):015004, 2019.

\bibitem{Visser:2004qp}
M.~Visser and S.~Weinfurtner.
\newblock {Massive phonon modes from a BEC-based analog model}.
\newblock {\em arXiv:cond-mat/0409639}, 2004.

\bibitem{Marino2019}
F.~Marino.
\newblock Massive phonons and gravitational dynamics in a photon-fluid model.
\newblock {\em Phys. Rev. A}, 100:063825, 2019.

\bibitem{JMMCthesis}
J.~M. Munoz-Castaneda.
\newblock {\em Efectos de borde en teoría cuántica de campos}.
\newblock {PhD thesis}, University of Zaragoza, 2009.

\bibitem{Galindobook}
A.~Galindo and P.~Pascual.
\newblock {\em Quantum Mechanincs (vols. I and II)}.
\newblock Springer, Heidelberg, Germany, 1991.

\bibitem{Gamelinbook}
T.~W. Gamelin.
\newblock {\em Complex Analysis}.
\newblock Springer, New York, USA, 2001.

\bibitem{DHN1974a}
R.~F. Dashen, B.~Hasslacher, and A.~Neveu.
\newblock {Nonperturbative methods and extended-hadron models in field theory.
  I. Semiclassical functional methods}.
\newblock {\em Phys. Rev. D}, 10:4114, 1974.

\bibitem{DHN1974b}
R.~F. Dashen, B.~Hasslacher, and A.~Neveu.
\newblock {Nonperturbative methods and extended-hadron models in field theory.
  II. Two-dimensional models and extended hadrons}.
\newblock {\em Phys. Rev. D}, 10:4130, 1974.

\bibitem{Kirstenbook}
K.~Kirsten.
\newblock {\em Spectral Functions in Mathematics and Physics}.
\newblock Chapman \& Hall/CRC, Florida, USA, 2001.

\bibitem{NISTbook}
F.~W. Olver, D.~W. Lozier, R.~F. Boisvert, and C.~W. Clark.
\newblock {\em NIST handbook of mathematical functions. 1st edn.}
\newblock Cambridge University Press, Cambridge, U.K., 2010.

\bibitem{Thirring1970}
W.~Thirring.
\newblock Systems with negative specific heat.
\newblock {\em Z. Phys. A - Hadrons and nuclei}, 235(4):339--352, 1970.

\bibitem{Cervero2005}
J.~M. Cerver\'o and A.~Rodr\'iguez.
\newblock {Simple model for a quantum wire III. Transmission in finite samples
  with correlated disorder}.
\newblock {\em Eur. Phys. J. B}, 43:543, 2005.

\bibitem{PhysRevD.17.2717}
C.~G. Callan, R.~Dashen, and D.~J. Gross.
\newblock Toward a theory of the strong interactions.
\newblock {\em Phys. Rev. D}, 17:2717--2763, 1978.

\bibitem{Mantonbook}
N.~Manton and P.~Sutcliffe.
\newblock {\em Topological Solitons}.
\newblock Cambridge University Press, Cambridge, U.K., 2004.

\end{thebibliography}

\end{document}